\documentclass[12pt]{article}
\usepackage{amsfonts}
\usepackage{latexsym,cite,amssymb}
\usepackage{times}
\usepackage{graphicx}
\usepackage{color}
\usepackage[english]{babel}

\makeatletter
\renewcommand\section{\@startsection {section}{1}{\z@}%
                                   {-3.5ex \@plus -1ex \@minus -.2ex}
                                   {2.3ex \@plus.2ex}%
                                   {\normalfont\large\bfseries}}

\renewcommand\subsection{\@startsection{subsection}{2}{\z@}%
                                     {-3.25ex\@plus -1ex \@minus -.2ex}%
                                     {1.5ex \@plus .2ex}%
                                     {\normalfont\bfseries}}

\renewcommand{\theequation}{\thesection.\arabic{equation}}

\makeatother

\newcommand{\bea}{\begin{eqnarray}}
\newcommand{\eea}{\end{eqnarray}}
\newcommand{\be}{\begin{equation}}
\newcommand{\ee}{\end{equation}}
\newcommand{\bem}{\begin{pmatrix}}
\newcommand{\eem}{\end{pmatrix}}
\newcommand{\bl}{\begin{align}}
\newcommand{\el}{\end{align}}

\def\S{\Sigma}

\begin{document}
\begin{flushright}
{\small MPP-2013-192 }
  \end{flushright}
\begin{center}
${}$ \thispagestyle{empty}

\vskip 1.5cm {\LARGE {\bf Striped phases in the holographic \\[2mm] insulator/superconductor transition}} \vskip 1.25 cm  {
Johanna Erdmenger ${}^{a}~$\footnote{ jke@mpp.mpg.de},
Xian-Hui Ge${}^{b,c}~$\footnote{ gexh@shu.edu.cn},
 ~~~ Da-Wei Pang${}^{a}~$\footnote{
dwpang@mpp.mpg.de}
 }
 ~~~
\vskip 0.5cm${}^{a}$Max-Planck-Institut f\"{u}r Physik (Werner-Heisenberg-Institut), F\"{o}hringer Ring 6, 80805
M\"{u}nchen, Germany\\
~{}
${}^{b}$Department of Physics, Shanghai University, ShangDa Road 99, 200444 Shanghai, China\\
${}^{c}$State Key Laboratory of Theoretical Physics, Institute of Theoretical Physics,\\
Chinese Academy of Sciences, Beijing 100190, China
~{}
\\

~~~\\
~~~\\

\vspace{1cm}

\begin{abstract}
\baselineskip=16pt
We study striped phases in holographic insulator/superconductor transition by considering a spatially modulated chemical potential
in the AdS soliton background. Generally striped phases can develop above a critical chemical potential. When the
constant leading term in the chemical potential is set to zero, a
discontinuity is observed in the  charge density as function of the chemical potential
in the limit of large wave vector. We explain this discontinuity using
an analytical approach. When the constant leading term in the chemical
potential is present, the critical chemical potential is larger than
in the case of a homogeneous chemical potential, which indicates that
the spatially modulated chemical potential disfavors the phase
transition. This behavior is again confirmed by an analytical approach. We also calculate the grand canonical potential and find that the striped phase is favored.

\end{abstract}
\end{center}

\newpage

\section{Introduction }

The AdS/CFT correspondence has provided a powerful framework for
investigating strongly coupled field theories via the corresponding
weakly coupled gravity duals~\cite{ads/cft,gkp,w}. In recent years
tremendous progress has been achieved
in applying generalized AdS/CFT models to systems of relevance for condensed matter
physics, high $T_{c}$ superconductors~\cite{Hartnoll:2008vx,
  Hartnoll:2008kx} for example. The simplest gravitational description
of high $T_{c}$ superconductivity is a black hole in
Einstein-Maxwell-charged scalar theory in AdS, where the
superconducting phase transition corresponds to an AdS black hole
forming scalar hair~\cite{gub1, gub2}.

High $T_{c}$ superconductors have a very rich phase
structure. In particular, close to the superconducting phase, there
exists an insulator phase with antiferromagnetic order, the Mott
insulator. High $T_{c}$ superconductivity may be implemented by
electronically doping the Mott insulator. A holographic version of this superconductor/insulator transition is
proposed in~\cite{takayanaki}, where the authors consider
Einstein-Maxwell-charged scalar theory in the five-dimensional AdS
soliton background. The AdS soliton metric is obtained by double Wick rotating the
five-dimensional Schwarzschild black hole,
\begin{equation}
\label{schads5}
ds^{2}=\frac{l^{2}dr^{2}}{f(r)}-r^{2}f(r)dt^{2}+r^{2}(dx^{2}+dy^{2}+dz^{2}),~~~~f(r)=r^{2}-\frac{r_{0}^{4}}{r^{4}},
\end{equation}
by substituting $t\rightarrow i\chi$ and $z\rightarrow it$. The resulting geometry reads
\begin{equation}
ds^{2}=\frac{l^{2}dr^{2}}{f(r)}+r^{2}(-dt^{2}+dx^{2}+dy^{2})+f(r)d\chi^{2},
\end{equation}
where $f(r)$ is still given by~(\ref{schads5}). The metric describes a cigar
with the tip at $r=r_{0}$. We need to impose periodicity $\chi\sim\chi+\pi l/r_{0}$ for the spatial coordinate $\chi$
to avoid a conical singularity at the tip. Note that the spacetime approaches $R^{1,2}\times S^{1}$ near the boundary and thus the
dual field theory lives in $2+1$ dimensions, according to AdS/CFT.

As pointed out in~\cite{takayanaki}, the AdS soliton background may be identified
as the insulator phase and the charged AdS black hole background is
identified with the superconducting phase.
The holographic insulator/superconductor transition is realized by
dialling the chemical potential. It turns out that the associated holographic
phase diagram displays qualitative similarity with the phase diagram
of the high $T_{c}$ cuprates. The holographic analysis just
described was performed in the probe limit. In a subsequent
paper~\cite{horowitz2}, the analysis of the phase diagram is completed
by including the backreaction. The phase structure with backreaction exhibits new features: For example, when lowering the temperature to zero at fixed chemical potential, the system becomes first a superconductor and then an insulator in a certain range of parameters.

These investigations were carried out for the case of
translational invariance in the spatial part in the gravity background.
However, for many properties of condensed matter systems, the lattice
structure plays a decisive role. This applies for instance to the Drude peak of the conductivity.
A further example is experimental evidence from neutron scattering, which indicates
that high-$T_c$ cuprates are not homogeneous, and doping plays a
vital role for the existence of the superconducting phase.
 Recent studies of inhomogeneities in weakly coupled superconductors
 \cite{martin} and the discovery of transport anomalies in
$La_{2-x}Ba_{x}CuO_4$, which are particularly prominent for $x=1/8$ \cite{li,berg}, strongly suggest that inhomogeneities may play an important role in high
$T_c$ superconductivity: The cuprates may be  ``striped''
superconductors. A striped phase is characterized by  doped charges
which are concentrated along
spontaneously generated domain walls between antiferromagnetic insulating regions \cite{emery}. Inhomogeneities arise since the electrons tend to cluster
in regions of suppressed antiferromagnetism. Experiments show that
strongly condensed ``static'' striped order may suppress superconductivity, but fluctuating striped order
might be beneficial to superconductivity \cite{tamada}.

Striped phases may be caused by charge density (CDW) or spin density waves (SDW). Signatures of CDW have been reported in a variety
of strongly correlated superconductors, such as $La_{1.6-x}Nd_{0.4x}Sr_x CuO_4$ and $La_{2-x}Ba_{x}CuO_4$. The CDW is described by a modulation of the charge density \cite{gruner}
\begin{equation}
\label{cdw15}
\rho(x)=\rho_0+\rho_{1}\cos(Qx+\theta),
\end{equation}
 where $\rho_0$ is the uniform charge density, $\rho_{1}$ is the
 amplitude of the CDW and $Q$ is the wave vector, and $\theta$ the
 phase of the condensate.

In contrast to the ground state of BCS superconductors, which consists of electron pairs,
 CDW are related to pairs of electrons and holes with parallel spins,
 while SDW are related to pairs of electrons and holes with opposite spins. Moreover, the CDW ground state is non-magnetic, but the SDW ground state has a well-defined
 magnetic character with associated low-lying magnetic
 excitations. The antiferromagnetic Mott insulator phase is found to be destroyed rapidly as holes are introduced by doping, and superconductivity appears.
To summarize, the CDW and SDW introduce instabilities related to the spontaneous breaking of the symmetries of the Euclidean group.

CDWs were studied within holography for different gravity backgrounds. Holographic CDWs with the desired property of spontaneous breaking of translational invariance were obtained in~\cite{Donos:2011bh}, where the charge density spontaneously acquires a spatially modulated vev.
For further holographic models with spontaneous breaking of
translational invariance, see~\cite{Donos:2011qt, Donos:2011ff,
  Donos:2012gg, Donos:2012wi, h4, h8, Donos:2013gda, h5,
  Withers:2013loa, Withers:2013kva}. Earlier work, for
instance~\cite{Flauger:2010tv}, considered CDWs in systems where
translational symmetry is explicitly broken by a spatially modulated
chemical potential. Since strictly speaking, a CDW requires
spontaneous symmetry breaking, these models may be viewed as toy
models of CDWs. Further work along these lines
includes~\cite{Hutasoit:2011rd, Ganguli:2012up, Hutasoit:2012ib}. -- A
further approach to holographic CDWs involving a two-form in the
gravity action is given in~\cite{aperis}.

The approach involving explicit breaking of translational invariance is also related to models of holographic lattices. These were realized for instance in~\cite{Horowitz:2012ky} by
introducing spatially modulated sources and numerically solving a set
of coupled PDEs. In the model of \cite{Horowitz:2012ky}, the optical conductivity exhibits a scaling behavior which matches the experiments very well. Subsequent generalizations in this direction include~\cite{Horowitz:2012gs, Horowitz:2013jaa, Ling:2013aya}.

This paper provides a first step toward a complete
holographic realization of the striped insulator/superconductor
transition in the presence of a spatially modulated chemical potential. Following~\cite{Flauger:2010tv}, we view this as a toy model for CDWs, though our breaking of translational invariance is explicit and not spontaneous. We leave the investigation of spontaneous breaking in the insulator/superconductor transition for future work. We consider Einstein-Maxwell-charged scalar theory in five-dimensional
AdS soliton background with a spatially modulated electrostatic potential $A_{t}=A_{t}(r,x)$, whose asymptotic behavior is given by
\begin{equation}
\label{Atrx}
A_{t}(r,x)\Big|_{r\rightarrow\infty}\sim\mu(x)-\frac{\rho(x)}{r^{2}},
\end{equation}
where $\mu(x)$ and $\rho(x)$ denote the chemical potential and charge
density respectively, both of which are spatially modulated. In addition, we split $A_{t}(r,x)$ into a homogeneous part $A_{0}(r)$ and an inhomogeneous part $A_{1}(r)\cos Qx$,
\begin{equation} \label{eq:At}
A_{t}(r,x)=A_{0}(r)+A_{1}(r)\cos Qx .
\end{equation}
 The asymptotic behavior of $A_{0}(r)$ and $A_{1}(r)$ is
\begin{equation}
\label{A0A1}
A_{0}(r)\Big|_{r\rightarrow\infty}\sim\mu_{0}-\frac{\rho_{0}}{r^{2}},~~
A_{1}(r)\Big|_{r\rightarrow\infty}\sim\mu_{1}-\frac{\rho_{1}}{r^{2}}.
\end{equation}
Then by combining~(\ref{Atrx}) and~(\ref{A0A1}), we obtain
\begin{equation}\label{mucharge}
\mu=\mu_{0}+\mu_{1}\cos Qx,~~~\rho=\rho_{0}+\rho_{1}\cos Qx,
\end{equation}
where the charge density takes a form similar to~(\ref{cdw15}) with $\theta=0$.

 We will work in the probe limit, such that the backreaction of the gauge field and the scalar field on the background may be neglected. Therefore the resulting equations of motion are ODEs, which simplifies both the numerical and analytical calculations considerably. Our main results are summarized as follows:

\begin{itemize}

\item We first consider the purely inhomogeneous case with $A_{0}=0$ in~(\ref{eq:At}), which corresponds to a single-mode CDW.
This case has  connections with experiments as many materials exhibit a single-mode CDW or one dominant wave-vector.
For simplicity, $\mu_{1}$ and $\rho_{1}$ will be referred to as the ``chemical potential'' and ``charge density'' for the purely inhomogeneous case. Our numerical analysis shows that the charge density $\rho_{1}$ exhibits a discontinuity as function of the chemical potential $\mu_{1}$, with a jump of size related to $Q$.
\item We calculate the condensate and the charge density analytically
  and find qualitative agreement with the numerical results. In
  particular, the discontinuity as described above is obtained analytically. The relation
  between the charge density $\rho_{1}$ and the chemical potential
  $\mu_{1}$ takes the form
\begin{equation}
   \rho_{1}\approx\lambda_{1}(\mu_{1}-\mu_{c})+\lambda_{2}Q^{2} \, ,
\end{equation}
    where $\lambda_{1}, \lambda_{2}$ are constants determined by
    analytical methods and $\mu_{c}$ denotes the critical chemical
    potential. When approaching the transition point,
    $\mu_{1}\rightarrow\mu_{c}$, we find that $\rho_{1}\sim Q^{2}$. As discussed in the main text below, such a discontinuity is known from condensed matter physics~\cite{gruner}.

\item The case $A_{0}\neq 0$ in~(\ref{eq:At}) for which homogeneous and
  inhomogeneous modes are mixed is also studied both numerically and
  analytically, where qualitative agreement is found once again. Comparing to the purely homogeneous case~\cite{takayanaki},
  we find that the critical value of the chemical potential receives minor corrections at small $Q$, while it becomes larger than in the purely homogeneous case, which means that the CDW impedes the phase transition at large wave vectors.
  Another interesting feature is that the homogeneous and
  inhomogeneous condensate compete with each other, such that the contribution $\mu_{1}$ to the critical chemical potential from the inhomogeneous part can even be negative. In this case, we have $\mu_{0}>|\mu_{1}|$, such that there is no discontinuity in the charge density.

\item The grand canonical potential is evaluated, which shows that the striped phase is favored.
\item The conductivity perpendicular to the direction of the stripes is
  calculated, which behaves in the same way as in the homogeneous case
  and does not receive corrections from the spatially modulated modes.
\end{itemize}

Before moving on, let us briefly review further literature on
holographic constructions of spatially modulated phases and
lattices. States breaking rotational and translational invariance in
holographic QCD were observed in~\cite{Domokos:2007kt}. The
holographic realization of spatially modulated unstable modes was
initiated in~\cite{ooguri,ooguri1}.
Holographic realizations  of spontaneously generated spatially modulated
phases in presence of a magnetic field may be found in
\cite{Ammon:2011je, Bu:2012mq, h9, h7}.  A holographic metal-insulator
transition transition in a helical lattice was given in \cite{Donos1}.

The new feature of the present work is to study spatial modulations for the soliton backgrounds dual to an insulating phase, and for the holographic insulator/superconductor transition.

The organization of this paper is as follows. In section 2, we explain
the basic setup for  holographic stripes and CDWs. In section 3, we solve the equation of motion with purely inhomogeneous
electrostatic potential $A_t(r)=A_1(r)\cos Qx$, where both numerical and analytical computations are performed. In section 4, we turn to more complicated situation of solving the equation of motion with
both homogeneous and inhomogeneous contributions present, i.e. $A_t(r,x)=A_0(r)+A_1(r)\cos Qx$.  The grand canonical potential is calculated in section 5 and the conductivity perpendicular to the direction of stripes is computed in section 6. Conclusions and
possible directions for future investigations are presented in section 7.

\section{The background}
\subsection{Charge density waves and the holographic duals}
We give a brief review on the essential physics of charge density waves~\cite{gruner} and their gravity duals.
It was pointed out by Peierls that for a one-dimensional metal coupled to the underlying lattice, the ground state is characterized by a collective
mode formed by electron-hole pairs with wave vector $Q=2k_{F}$. The charge density of the collective mode is
\begin{equation}
\rho(\vec{r})=\rho_{0}+\rho_{1}\cos(2\vec{k}_{F}\cdot\vec{r}+\varphi),
\end{equation}
where $\rho_{0}$ denotes the unperturbed electron density of the metal. The condensate is referred to as the charge density wave (CDW).
The order parameter is complex,
\begin{equation}
\Delta=|\Delta|e^{i\varphi}.
\end{equation}
Translational symmetry is broken for CDW ground states and the
collective excitations are referred to as
phasons and amplitudons, which correspond to fluctuations of the phase and amplitude of the condensate.

Motivated by this condensed matter picture, a holographic model of CDW was proposed in~\cite{aperis}, which consists of the modulus and phase of
a complex scalar field, a $U(1)$ gauge field and an antisymmetric field. Signatures of the CDW can be observed by studying the collective modes and
the dynamical response to an external electric perturbation. In the
model of~\cite{aperis}, both the charge density and the chemical potential are determined by numerically solving the relevant equations of motion. Moreover, a single-mode CDW is considered. Also in condensed matter physics, many materials exhibit either a single-mode CDW or only one dominant wave vector.

On the other hand, within condensed matter physics, there are models
based on the coexistence of homogeneous superconductivity with
CDWs. This led the authors of~\cite{Flauger:2010tv} to construct a
corresponding holographic model. The bulk theory is 3+1-dimensional
Einstein-Maxwell-scalar theory and the CDW is sourced by a modulated
chemical potential. The main focus of ~\cite{Flauger:2010tv} is to
study the interactions between superconductivity and CDWs, hence the
CDW is chosen to be sourced by a modulated chemical potential, which
explicitly breaks translation invariance.

As mentioned in the
introduction, subsequent models realize the desired property of a
dynamically generated CDW, for instance~\cite{Donos:2011bh, Donos:2013gda}.
Here, for considering the insulator/superconductor transition,  we
follow the approach of~\cite{Flauger:2010tv} for simplicity.

\subsection{Basic setup}
Let us consider the five-dimensional Einstein-Maxwell theory with a charged scalar field

\be
S=\int d^5x\sqrt{-g}\bigg(R+\frac{12}{l^2}-\frac{1}{4}F^{\mu\nu}F_{\mu\nu}-|\partial_{\mu}\Psi-iqA_{\mu}\Psi|^2
-m^2|\Psi|^2\bigg),
\ee
 where the cosmological
constant is $\Lambda=-6/l^2$ and
$F_{\mu\nu}=\partial_{\mu}A_{\nu}-\partial_{\nu}A_{\mu}$.
The equation of motion for the charged scalar is
\be
-\frac{1}{\sqrt{-g}}D_{\mu}\bigg(\sqrt{-g}g^{\mu\nu}D_{\nu}\Psi\bigg)+m^2 \Psi=0.
\ee
The Maxwell field equation reads
\be
\frac{1}{\sqrt{-g}}\partial_{\mu}\bigg(\sqrt{-g}F^{\mu\nu}\bigg)=iq g^{\mu\nu}\bigg[\Psi^{*}D_{\nu}\Psi-\Psi(D_{\nu}\Psi)^{*}\bigg].
\ee
The Einstein's equations are given by
 \be\label{einstein}
 R_{\mu\nu}-\frac{1}{2}g_{\mu\nu}R-\frac{6}{l^2}g_{\mu\nu}= \frac{1}{2}T_{\mu\nu},
 \ee
 where \be
 T_{\mu\nu}=F_{\mu\lambda}F^{\lambda}_{\nu}-\frac{1}{4}g_{\mu\nu}F^{\lambda \rho}F_{\lambda \rho}
 -g_{\mu\nu}(|D \Psi|^2+m^2|\Psi|^2)+\bigg[D_{\mu} \Psi (D_{\nu} \Psi)^{*}+D_{\nu} \Psi (D_{\mu} \Psi)^{*}\bigg].
 \ee

For holographically describing the insulator phase we use the
the five-dimensional planar AdS soliton, whose metric is given by \cite{witten,horowitz1}
\bea
&&ds^2=\frac{l^2 dr^2}{f(r)}+r^2(dx^2+dy^2-dt^2)+f(r)d\chi^2,\\
&&f(r)=r^2-\frac{r^4_{0}}{r^2}.
\eea
Note that the AdS soliton solution may be derived by a double Wick rotation of the AdS Schwarzschild black hole, and $\chi$ should be identified as
$\chi\sim\chi+\pi l/r_{0}$ to ensure a smooth geometry. The resulting metric describes a cigar
with the tip at $r=r_{0}$.

We work in the probe limit and take into account the coupling of this system to the inhomogeneous gauge field. The backreaction of the gauge field and scalar to the background geometry will be neglected.
We consider a non-zero electrostatic potential of the form
\be
A_t=A_t(r,x).
\ee
After performing a coordinate transformation $z=\frac{r_{0}}{r}$ and
setting $r_{0}=l=1$, the metric takes the form
\bea
&&ds^2=\frac{1}{z^2 h(z)}dz^2+\frac{1}{z^2}\bigg(-dt^2+dx^2+dy^2\bigg)+\frac{1}{z^2}h(z)d\chi^2,\\
&&h(z)=1-z^4.
\eea
The equations of motion are then given by
\bea
&&\Psi''+(\frac{h'}{h}-\frac{3}{z})\Psi'+\frac{q^2 A^2_t}{h}\Psi-\frac{m^2\Psi}{z^2h}+\frac{\partial^2_x\Psi}{h }=0,\label{main1}\\
&&A''_t+(\frac{h'}{h}-\frac{1}{z})A'_t-\frac{2\Psi^2}{z^2h}A_t+\frac{\partial^2_x A_t}{h }=0,\label{main2}
\eea
where the prime $'$ denotes a derivative with respect to $z$. We focus on the case where $m^2=-\frac{15}{4}$, which stays above the BF bound given by $m_{BF}^2=-\frac{(D-1)^2}{4}=-4$.
The effective mass term for the scalar field is given by
\be
  m^2_{\rm eff}=m^2-g^{xx}\partial^2_x\Psi/\Psi +g^{tt}q^2A^2_t.
\ee
Because $g^{tt}$ is negative outside the tip at $z=1$, the effective mass $m^2_{\rm eff}$ can become negative when the terms $-g^{xx}\partial^2_x\Psi/\Psi +g^{tt}q^2A^2_t$ take negative values.
We will see in the below that the term  $-g^{xx}\partial^2_x\Psi/\Psi$
is positive and may impede the formation of a condensate.

The asymptotic boundary behavior of the fields is given by
\bea
&&\Psi(z\rightarrow 0)=\Psi^{(1)}(x){z^{3/2}}+\Psi^{(2)}(x){z^{5/2}}+...,\\
&&A_t(z\rightarrow 0)=\mu(x)-\rho(x) z^2+...,
\eea
where $\mu$ and $\rho$ are the chemical potential and the charge density in the dual field theory, respectively.
The constants $\Psi^{(1)}$ and $\Psi^{(2)}$ are both normalizable and may be used to define operators in the dual field theory with mass dimension $\Delta=3/2$ and $\Delta=5/2$, respectively.

In order to study the effect of an inhomogeneity in this strongly coupled system, we consider a modulated electrostatic potential of the form
\be
\label{Atzx}
A_t (z,x)=A_{0}(z)+  A_{1}(z)\cos Qx,
\ee
where $Q$ is the wave number along the $x$-direction.
As discussed in the introduction, this ansatz may be interpreted as a CDW. The asymptotic behavior of $A_{0}(z)$ and $A_{1}(z)$ are given by
\begin{equation}
\label{A0A1z}
A_{0}(z)(z\rightarrow0)\sim\mu_{0}-\rho_{0}z^{2},~~
A_{1}(z)(z\rightarrow0)\sim\mu_{1}-\rho_{1}z^{2}.
\end{equation}
Then by combining~(\ref{Atzx}) and~(\ref{A0A1z}), we obtain
\begin{equation}
\mu=\mu_{0}+\mu_{1}\cos Qx,~~~\rho=\rho_{0}+\rho_{1}\cos Qx,
\end{equation}When $A_1(z)=0$, the inhomogeneity disappears. Near the boundary, the homogeneous part of $A_t (z,x)$ is simply given by
\be
A_{0}(z)=\mu_{0}.
\ee
Note that near the tip $z=1$, the fields behave as
\bea
&&\Psi(z)=a+b \log(1-z)+c(z-1)+...\label{neu1}\\
&&A_{0}(z)=A+B\log(1-z)+C (z-1)+...\label{neu2}
\eea
We impose a Neumann boundary condition, which requires $b=B=0$. For the scalar field, we work in Fourier space and expand
\be
\Psi(x,z)=\sum^{\infty}_{n=0}\psi_n(z)\cos(nQx),
\ee
which is an even function for $n$ and thus $\psi_{-n}=\psi_n$.
According to the AdS/CFT dictionary, the operator $\mathcal{O}$ dual to the scalar field $\Psi$ is
\be
\label{2eq33}
\langle\mathcal{O}^i\rangle=\sum^{\infty}_{n=0}\langle\mathcal{O}^{i}_n\rangle\cos nQx, ~~i=1,2,
\ee
with
\be
\label{2eq34}
\langle\mathcal{O}^i_n\rangle=\psi_n(z=0).
\ee
Up to the $n$-th order, the equations of motion for the modes of the scalar field are given by
\begin{eqnarray}
&&\psi''_0+(\frac{h'}{h}-\frac{3}{z})\psi'_0-\frac{m^2 \psi_0}{z^2 h}
+\frac{q^2}{h}\bigg((A^2_0+\frac{1}{2}A_1^2)\psi_0+A_0 A_1 \psi_1+\frac{A^2_1}{4}\psi_2\bigg)=0,\label{sec2psi0}\\
&&\psi''_1+(\frac{h'}{h}-\frac{3}{z})\psi'_1-\frac{m^2 \psi_1}{z^2 h}-\frac{Q^2 \psi_1}{ h}\\
& &+\frac{q^2}{h}\bigg((A^2_0+\frac{1}{2}A_1^2) \psi_1+2A_0 A_1 \psi_0+A_0A_1\psi_2+\frac{A^2_1}{4}(\psi_1+\psi_3)\bigg)=0,\label{sec2psi1}\nonumber\\
&&~~~~~~~~~~~~~~~~~ \vdots\nonumber\\
& &\psi''_n+(\frac{h'}{h}-\frac{3}{z})\psi'_n-\frac{m^2 \psi_n}{z^2 h}-\frac{Q^2 n^2 \psi_n}{h}\\
& &+\frac{q^2}{h}\bigg[(A^2_0+\frac{1}{2}A^2_1)\psi_n+A_0 A_1(\psi_{(n-1)}+\psi_{(n+1)})+
\frac{1}{4}A^2_1(\psi_{(n-2)}+\psi_{(n+2)})\bigg]=0.\nonumber
\end{eqnarray}
In the above equation for $\psi_n$, there are also other relevant modes $\psi_{(n-1)}$, $\psi_{(n+1)}$, $\psi_{(n-2)}$ and $\psi_{(n+2)}$. These terms mix with the $\psi_n$ terms such that solving this system becomes extremely involved. On the other hand, it has been shown in~\cite{Flauger:2010tv} that higher modes ($n>1$) are significantly suppressed in the Schwarzschild-${\rm AdS}_{4}$ black hole background. We expect that similar behavior is also present in our case. As an approximation, we therefore restrict our attention to the zeroth and first order of $\psi_n$ in the computation and set all higher modes $(n>1)$ to zero. A similar strategy is also followed in the condensed matter paper~\cite{gruner}.
Note that there is a scaling symmetry in the equations of motion
\be
\psi_n \rightarrow \lambda \psi_n,~~~A_t\rightarrow \lambda A_t,~~~\mu\rightarrow \lambda \mu,~~~q\rightarrow q/\lambda.
\ee
The probe limit corresponds to $\lambda \ll 1$, that is to say, $\mu \ll 1$ and $q \gg 1$ but $\mu q$ is kept finite. In the following we simply choose $q=1$.
In the normal phase, the $z$-dependent part of the equation of motion for $A_{1}$ is given by
\be \label{A1}
A''_1+\bigg(\frac{h'}{h}-\frac{1}{z}\bigg)A'_1-\frac{Q^2 A_1}{h}=0,
\ee
with Neumann-like boundary condition
\be
A_1(1,x)=1,~~~A_1(0,x)=\mu_{1}.
\ee
\subsection{Possible approximation method}

It is very difficult to obtain the exact solution of (\ref{A1}) and an
appropriate approximation method would be very welcome. Here we
investigate whether a method presented in \cite{Flauger:2010tv} is
applicable to our model. Unfortunately, the answer is negative.
Nevertheless, since this analysis is instructive, we present it here.

For Schwarzschild-$\textrm{AdS}_{4}$ dual to a $(2+1)$-dimensional gauge theory, it was shown
in~\cite{Flauger:2010tv} that the solution to eq.(\ref{A1}) can be well-approximated by solving a simplified version of the corresponding equation. The equation for $A_{1}$ in~\cite{Flauger:2010tv} is given by
\begin{equation}
\label{A1fl}
(1-z^{3})A_{1}^{\prime\prime}=Q^{2}A_{1}
\end{equation}
and the approximate solution was obtained analytically by solving the equation which can be obtained by taking the limit $z\rightarrow0$ in~(\ref{A1fl}). The approximate solution and the numerical solution to~(\ref{A1fl}) were compared and it was found that the agreement is excellent both for $Q\ll1$ and $Q\gg1$, and still reasonably good for
$Q\simeq \mathcal{O}(1)$.

Can we solve~(\ref{A1}) in a similar way given that our gravity theory is now five-dimensional? Let us consider the following simplified equation
\be\label{besel}
A''_1-\frac{1}{z}A'_1-{Q^2 A_1}=0,
\ee
which is obtained by taking $z\rightarrow0$ in~(\ref{A1}). The solution of~(\ref{besel}) may be obtained analytically and is given by
\be\label{analy}
A_1=\frac{1}{J_1(iQ)}\bigg[2 z J_1(iQz)-i\pi Q z J_1(i Q z)Y_1(-iQ)+i\pi Q z J_1(i Q)Y_1(-iQz)\bigg],
\ee
where $J_1(iQz)$ and $Y_1(iQz)$ denote Bessel functions of the first and the second kind, respectively.
\begin{figure}[htbp]
 \begin{minipage}{1\hsize}
\begin{center}
\includegraphics*[scale=0.59] {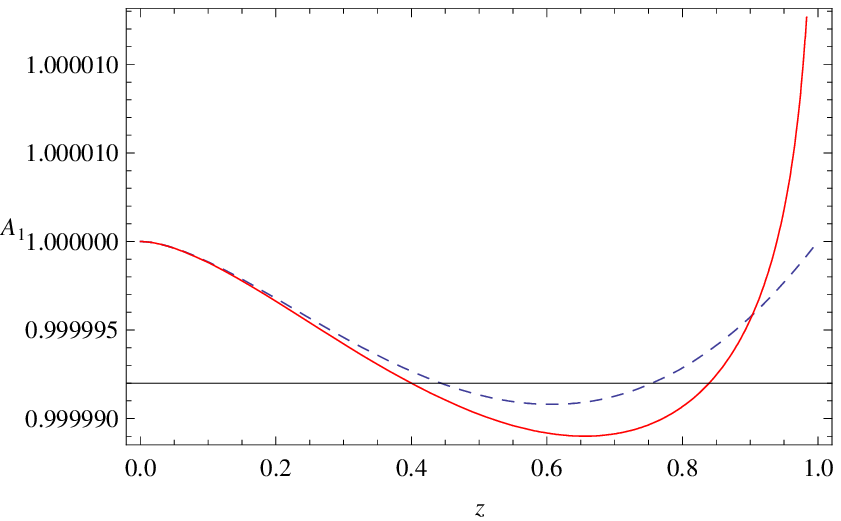}
\includegraphics*[scale=0.55] {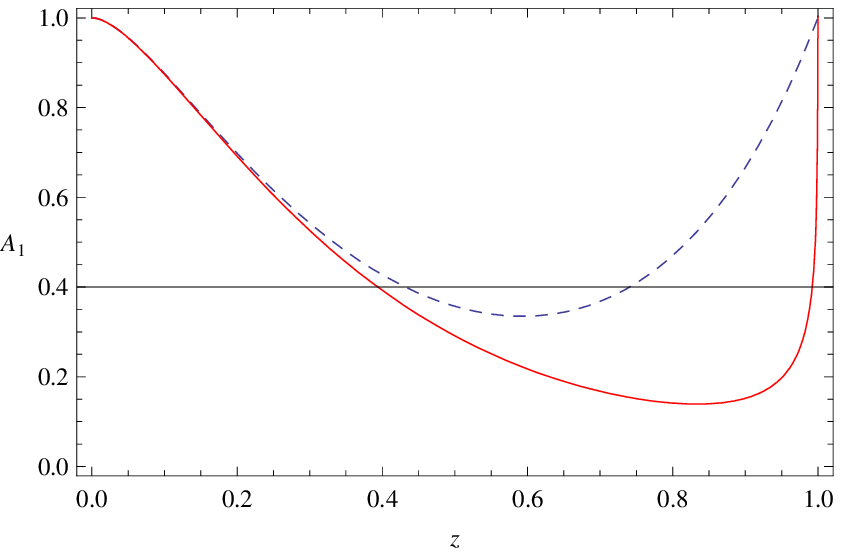}
\includegraphics*[scale=0.55] {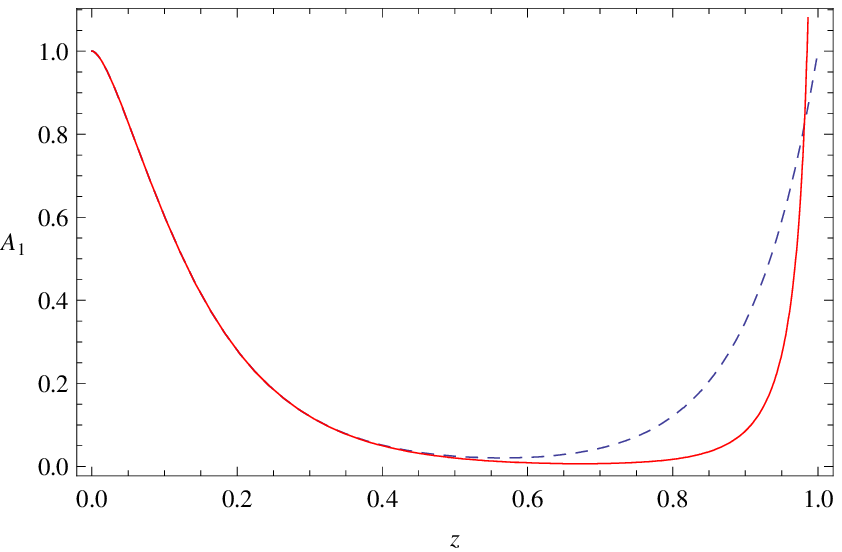}
\end{center}
\caption{(color online) The numerical exact (red and solid line) and analytically approximate (blue and dashed line) solutions of (\ref{A1}) for
various values of Q, with $Q=0.01$ (left), $Q=4$ (middle) and $Q=10$ (right).} \label{fig1}
\end{minipage}
\end{figure}
We plot the behavior of the exact numerical solution of (\ref{A1}) and the approximate solution (\ref{besel}) for various values of $Q$
in Fig.\ref{fig1}. It may be seen that unlike the case discussed in~\cite{Flauger:2010tv}, these two solutions do not match so well as expected. The agreement between these two solutions is quantitative acceptable only for $Q\ll 1$ and $Q\gg 1$, while there is significant disagreement for $Q=\mathcal{O}(1)$.

This means that the analytical solution (\ref{analy}) of (\ref{besel})
cannot be used to approximate the exact numerical solution of
(\ref{A1}), which is quite different from the four-dimensional black
hole cases discussed in~\cite{Flauger:2010tv}. The method appears to
be working well for approximating the solution in special cases, but may not be generalized to higher dimensions.

\section{Pure inhomogeneous solutions for $A_0(z)= 0$}

For explicit results, let us begin by considering a simpler case where the
electrostatic potential just contains the inhomogeneous part,
$A_t=A_{1}(z)\cos Qx$. Recall that the asymptotic behavior of $A_{1}$ is $A_{1}\sim\mu_{1}(x)-\rho_{1}(x)z^{2}$, which corresponds a single mode CDW in the dual boundary theory with a charge density of the form $\rho_{1}(x)\sim \cos Qx$, as considered in~\cite{aperis}. Moreover, the resulting model is relatively easily tractable numerically
and can provide direct insight into the phase transition. Note that although the actual charge density
and chemical potential are given by $\rho_{1}\cos Qx$ and $\mu_{1}\cos Qx$, we will refer to $\rho_{1}$ and $\mu_{1}$ as the ``charge density'' and ``chemical potential'' for simplicity. We first present our numerical computation. Then we will solve the equation of motion by using the Sturm-Liouville eigenvalue method. As we will see, the numerics and the analytical results match with each other.
\subsection{Numerics}
For the pure inhomogeneous case $A_{0}(z)=0$, the boundary conditions at the horizon are still of Neumann type and we set $A_n(0)=0$ ($1<n<n_{max}$) as in~\cite{Flauger:2010tv}.
We solve the following equations of motion for $A_1$ and $\psi_0$ numerically,
\bea
&&A''_1+\bigg(\frac{h'}{h}-\frac{1}{z}\bigg)A'_1-\frac{2\psi^2_0}{z^2 h}A_1-\frac{Q^2 A_1}{h}=0,\label{At}\\
&&\psi''_0+(\frac{h'}{h}-\frac{3}{z})\psi'_0-\frac{m^2 \psi_0}{z^2 h}
+\frac{1}{2h}A^2_1\psi_0=0,\label{scalar}\\
&&~~~~~~~~~~~~~~~~~ \vdots \nonumber\\
&&\psi''_n+(\frac{h'}{h}-\frac{3}{z})\psi'_n-\frac{m^2 \psi_n}{z^2 h}-\frac{Q^2 n^2\psi_n}{ h}\nonumber\\
&&+\frac{1}{h}\bigg[\frac{ A^2_1}{2}(\psi_{(n)}+\frac{1}{2}\psi_{(n-2)}
+\frac{1}{2}\psi_{(n+2)})\bigg]=0.\label{pn}
\eea
 We find that the numerical results are characterized by some typical values of the wave number $Q$
 and we mainly take $Q=0.01$ and $Q=1$ as two concrete examples in the following.
\begin{figure}[htbp]
 \begin{minipage}{1\hsize}
\begin{center}
\label{fig2}
\includegraphics*[scale=0.6] {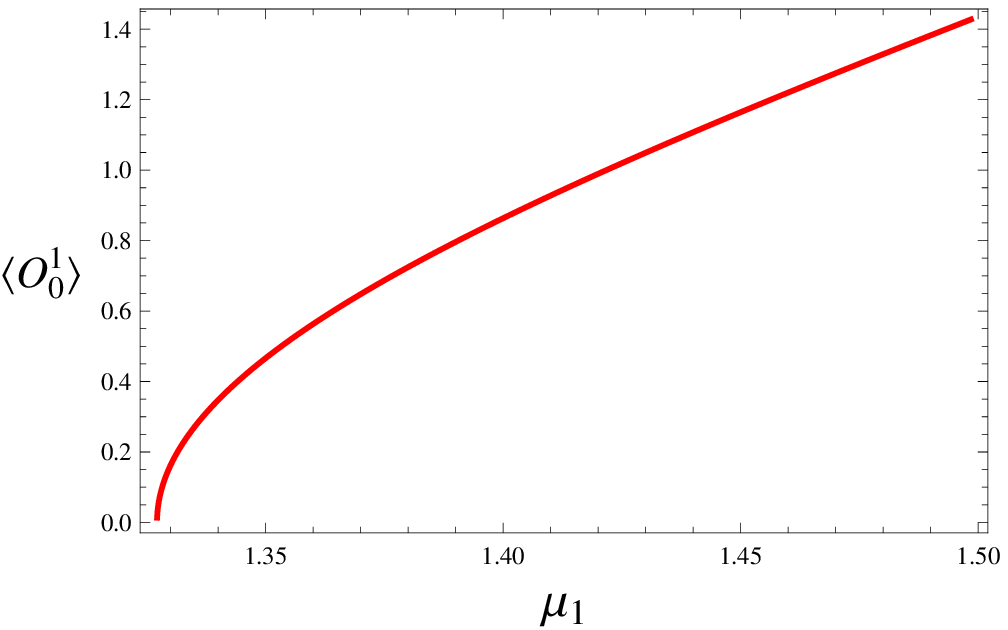}
\includegraphics*[scale=0.6] {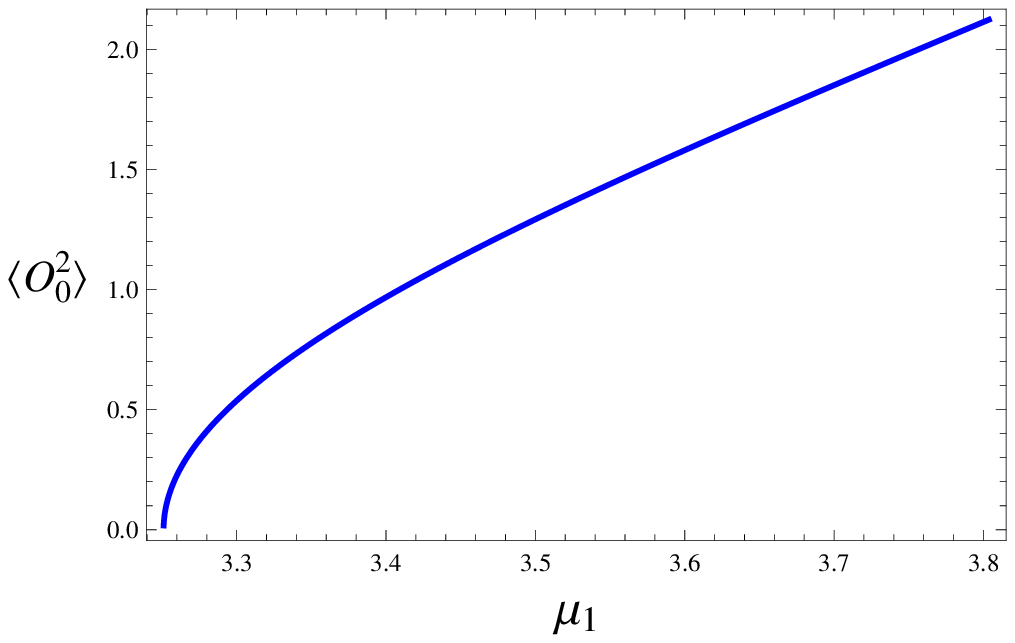}
\end{center}
\caption{(color online) The condensates of the scalar operators $\langle\mathcal{O}^{1}_0\rangle$ (left) and $\langle\mathcal{O}^{2}_0\rangle$ (right) for $Q=1$. $\langle\mathcal{O}^{1}_0\rangle$ and $\langle\mathcal{O}^{2}_0\rangle$ are defined in~(\ref{2eq33}) and~(\ref{2eq34}).} \label{condensate1}
\end{minipage}
\end{figure}

$\bullet ~~~\textsl{Q=1}~case.$~~~  We plot $\langle\mathcal{O}^1_{0}\rangle$ and $\langle\mathcal{O}^2_{0}\rangle$ as functions of the chemical potential $\mu_{1}$ in Fig.\ref{condensate1} for mass dimension $\Delta=3/2$ and $\Delta=5/2$, respectively. We find that the condensation occurs for $\mu_1 >1.33$ (left) and $\mu_1 >3.25$ (right).
\begin{figure}[htbp]
 \begin{minipage}{1\hsize}
\begin{center}
\includegraphics*[scale=0.61] {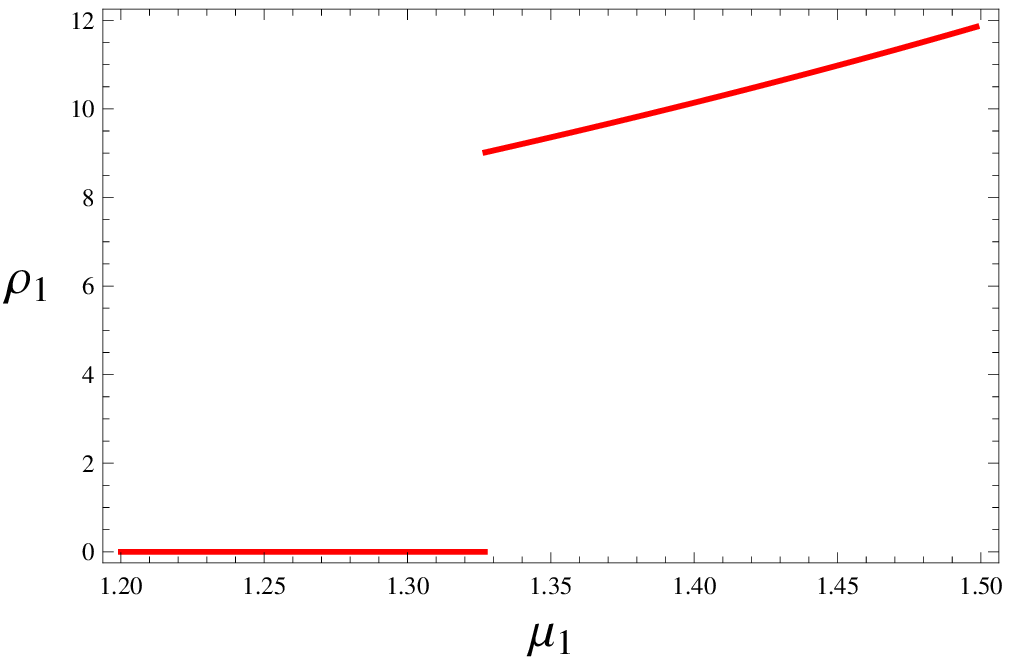}
\includegraphics*[scale=0.6] {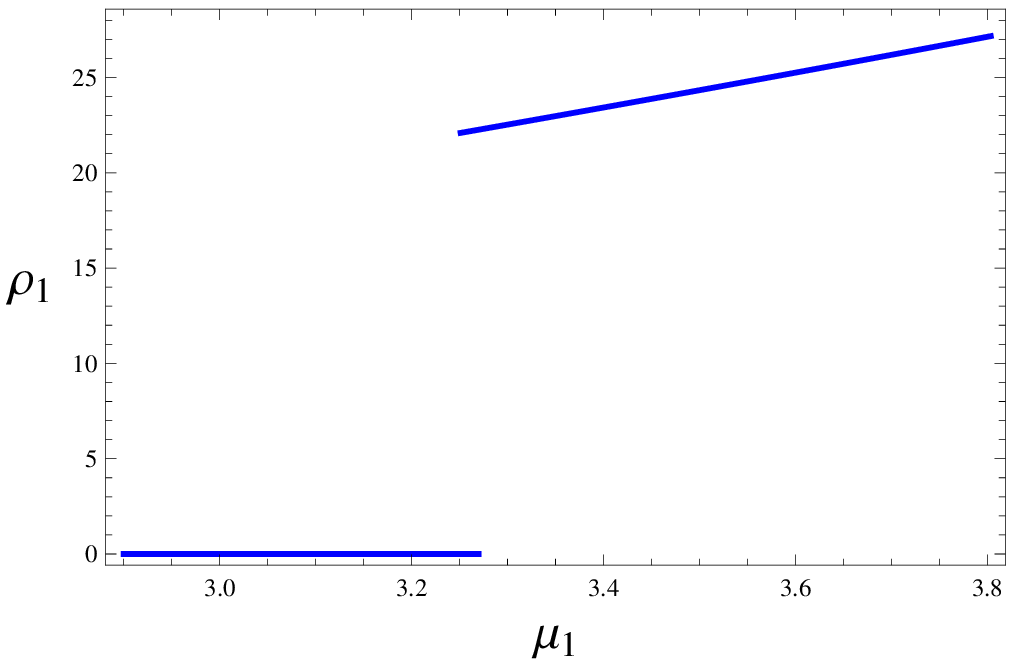}
\end{center}
\caption{(color online) The charge density $\rho_{1}$ of~(\ref{A0A1z}) plotted as a function of $\mu_{1}$ with $Q=1$.} \label{chargedensity3}
\end{minipage}
\end{figure}
For $Q=1$, we also plot the charge density $\rho_{1}$ as a function of the chemical potential $\mu_{1}$ in Fig.\ref{chargedensity3}. Interestingly, there exists a discontinuity in the charge density curve, which may be related to the effect of the wave number $Q$. A similar
discontinuity was also observed in~\cite{horowitz2}, where the backreaction of the Maxwell field and the charged scalar to the metric were considered without CDW. There the charge $q$ of the scalar field varies and when $q<1.2$, plots of the charge density versus chemical potential also exhibit a discontinuity. The reason for the discontinuity was unclear in~\cite{horowitz2}, while for our case, at least for
pure inhomogeneous $A_{t}$ in the probe limit, the discontinuity may
be attributed to the spatial modulation. To confirm this argument, we will also consider the small $Q$ limit.

\begin{figure}[htbp]
 \begin{minipage}{1\hsize}
\begin{center}
\includegraphics*[scale=0.6] {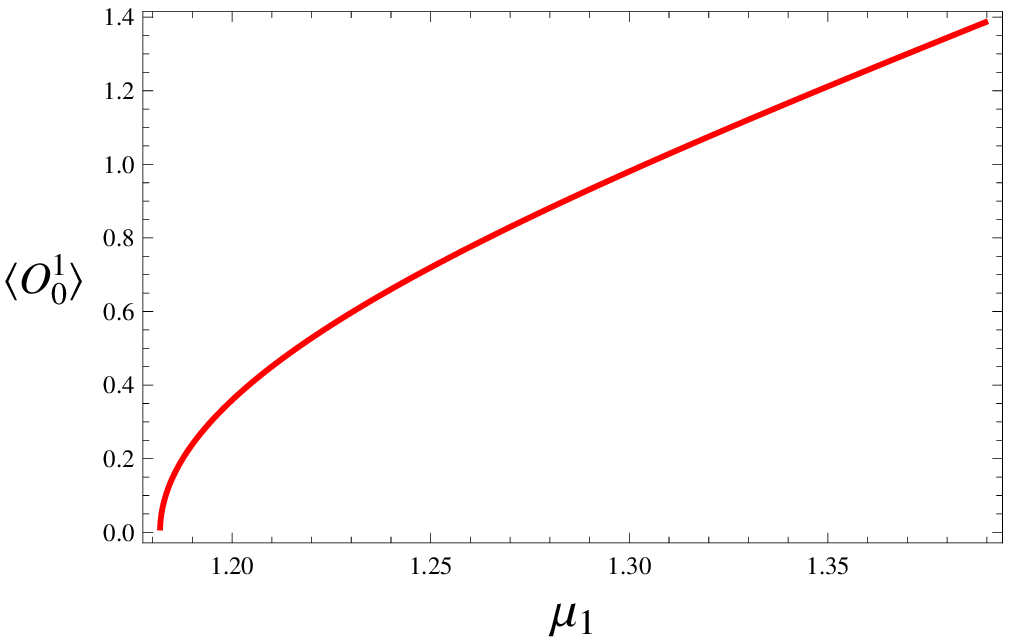}
\includegraphics*[scale=0.6] {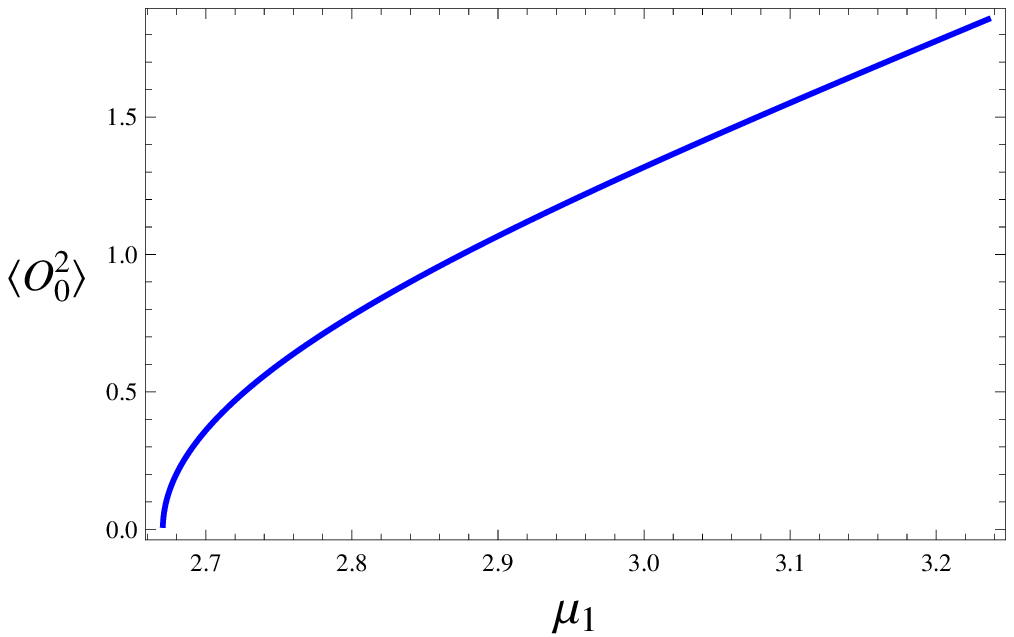}
\end{center}
\caption{(color online) The value of the condensate $\langle\mathcal{O}^{1}_0\rangle$ (left) and $\langle\mathcal{O}^{2}_0\rangle$ (right) as a function of chemical potential at the value $Q=0.01$.} \label{condensate8}
\end{minipage}
\end{figure}

\begin{figure}[htbp]
 \begin{minipage}{1\hsize}
\begin{center}
\includegraphics*[scale=0.6] {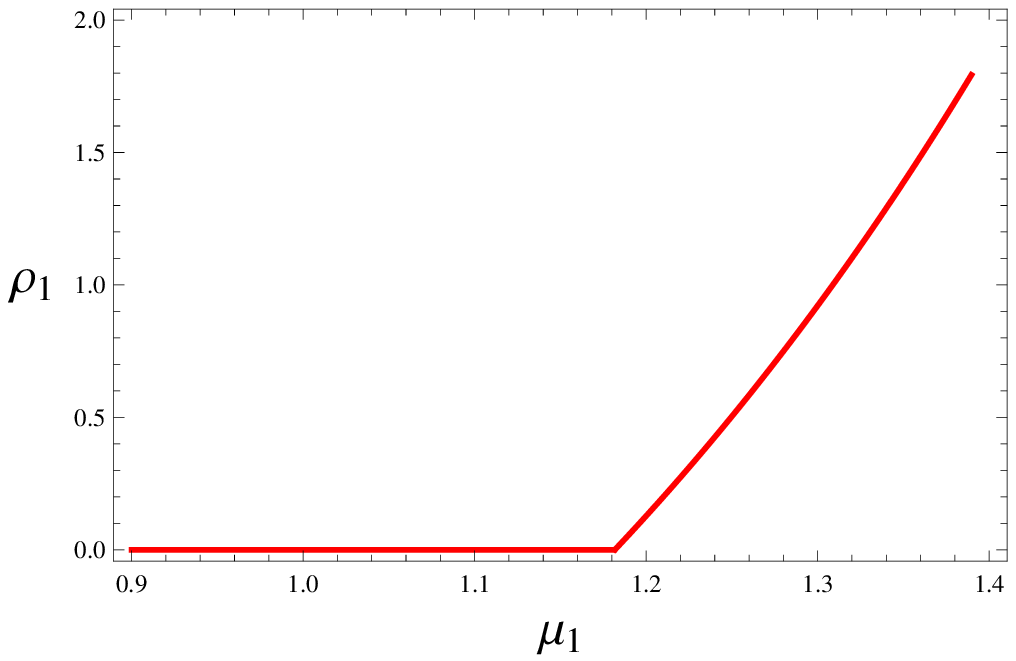}
\includegraphics*[scale=0.6] {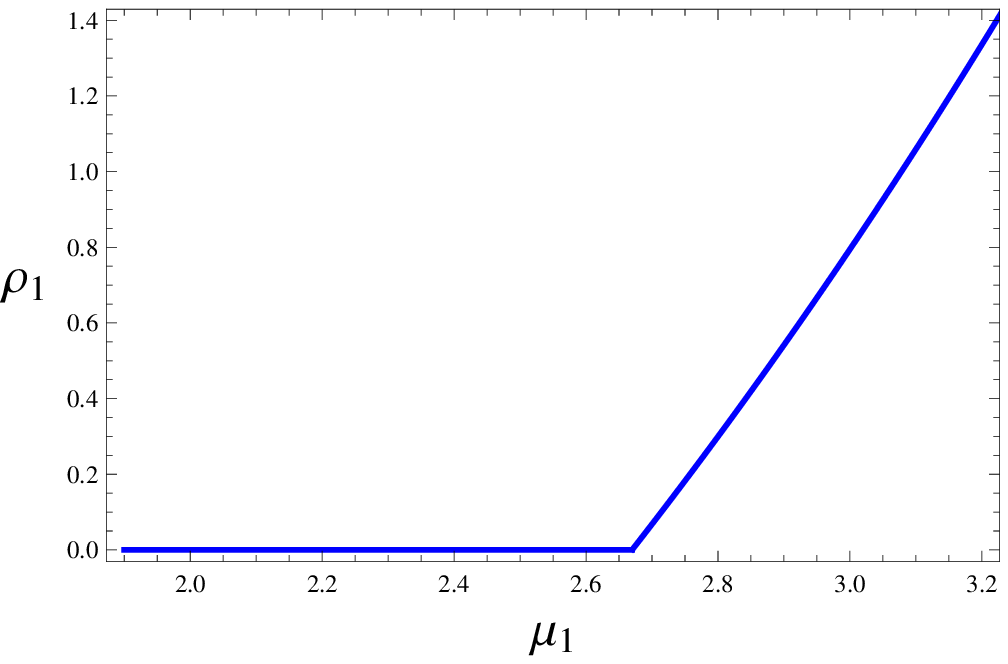}
\end{center}
\caption{(color online) The charge density $\rho_{1}$ plotted as a function of $\mu_{1}$.} \label{chargedensity}
\end{minipage}
\end{figure}

$\bullet ~~~\textsl{Q=0.01}~case.$~~~For $Q=0.01$ case, Fig.\ref{condensate8} demonstrates that the condensation occurs at $\mu_1 > 1.18$ and $\mu_1 > 2.68$ for mass
 dimension $\Delta=3/2$ and $\Delta=5/2$, respectively. Fig.\ref{chargedensity} shows that when $\mu_1$ is small,
  the system is described by the AdS soliton, which is interpreted as the insulating phase. As $\mu_1$ increases, the system reaches a superconducting phase. Moreover, the discontinuity seems to disappear in the small $Q$ limit. We
  will explain this phenomenon through analytical methods.

\subsection{Analytical calculation}
We are now going to solve the equations (\ref{At}-\ref{pn}) by using the Sturm-Liouville  eigenvalue method first developed in~\cite{siopsis}. This method was applied to the insulator/superconductor transition in \cite{cai1},
and was extended to various other conditions\cite{pan,cai2,pan2,pan20,lee,zhao,gang,wang}.

{$\bullet$ \textsl{Operator dimension $\Delta=\frac{3}{2}$}.}\\
We first consider the operator $\mathcal{O}_0^{1}$ of conformal dimension $\Delta=3/2$ and choose $m^2=-\frac{15}{4}$ as in \cite{takayanaki}.
As the chemical potential approaches the critical value, equation (\ref{scalar}) becomes
\be
\psi''_0+(\frac{h'}{h}-\frac{3}{z})\psi'_0+\frac{15 \psi_0}{4 z^2 h}
+\frac{1}{2h}\mu^2_1\psi_0=0.
\ee
In order to solve this equation by the Sturm-Liouville eigenvalue method, we need introduce a trial function $F(z)$ as
\be
\psi_0=\langle\mathcal{O}_0^{1}\rangle z^{3/2}F(z).
\ee
We then obtain
\be
F''+\frac{4z^3}{z^4-1}F'+\frac{-9z^4+2\mu^2_1z^2}{4z^2(1-z^4)}F=0.
\ee
The above equation can be recast as
\be
[(z^4-1)F']'+\frac{9}{4}z^2F-\frac{1}{2}\mu^2_1 F=0.
\ee
By using the Sturm-Liouville eigenvalue problem method, we write down the expression which can be used to estimate
the minimum eigenvalue of $\mu^2_1$
\be
\mu^2_1=\frac{\int^1_0dz (p F'^2+q F^2)}{\int^1_0dz s F^2},
\ee
with
\bea
p=z^4-1, q=-\frac{9}{4}z^2, s=-\frac{1}{2},
\eea
and the trial function $F(z)=1-\alpha z^2$.
We finally find the minimum value
\be
\mu_{min}\simeq 1.39
\ee
when $\alpha=0.230$. The critical value $\mu_{c}$ corresponds to $\mu_{min}$ and   thus in close agreement with the numerical value $\mu_{c}=1.33$ found in previous subsection.

When the chemical potential is above $\mu_c$, we can recast (\ref{At}) in terms of the scalar field as
\be
A''_1+\bigg(\frac{h'}{h}-\frac{1}{z}\bigg)A'_1-\frac{2\langle\mathcal{O}_0^{1}\rangle^2 z F^2}{h}A_1-\frac{Q^2 A_1}{h}=0,\label{Amain}
\ee
Near the critical point, $\langle\mathcal{O}_0^{1}\rangle$ is small and can serve as an expansion parameter. We would like to expand $A_1$ in series of $\langle\mathcal{O}_0^{1}\rangle$ as
\be
A_1 \sim \mu_c+\langle\mathcal{O}_0^{1}\rangle \chi(z)+...
\ee
Note that $\chi(z)$ obeys the boundary condition $\chi(1)=0$ at $z=1$. We obtain the equation of motion for $\chi(z)$ as
\be
\label{3eq56}
\chi''-\frac{1+3z^4}{z-z^5}\chi'=\frac{2\langle\mathcal{O}_0^{1}\rangle z\mu_cF^2}{h}+\frac{Q^2}{\langle\mathcal{O}_0^{1}\rangle h}\mu_c.\label{chi}
\ee
The above equation can be solved in the region $z\rightarrow 0$. One can easily find that near $z=0$, the scalar potential $A_1$ acts as
\be
\label{3eq57}
A_1\sim \mu_{1}-\rho_{1} z^2\simeq \mu_c+\langle\mathcal{O}_0^{1}\rangle \bigg(\chi(0)+\chi'(0)z+\frac{1}{2}\chi''(0)z^2+...\bigg).\label{41}
\ee
At zeroth order, we have
\be
\mu_{1}-\mu_c\simeq \langle\mathcal{O}_0^{1}\rangle\chi(0). \label{mu}
\ee
Comparing the  $z^1$ term on both sides of~(\ref{3eq57}), we obtain $\chi'(0)=0$.
Integrating~(\ref{3eq56}) we obtain
\bea
\chi(z)\bigg|^1_0&=&-2 \langle\mathcal{O}_0^{1}\rangle \mu_c \int^1_0\frac{z}{z^4-1}\bigg(\int^1_z F^2(x)dx\bigg)dz\nonumber\\&-&\int^1_0\bigg(\int^1_z\frac{Q^2\mu_c}{\langle\mathcal{O}^{1}_0\rangle x}dx\bigg)\frac{z}{z^4-1}dz.
\eea
At the boundary, we have
\bea
\chi(0)&=&\frac{\langle\mathcal{O}_0^{1}\rangle \mu_c}{60}  \left(\alpha^2 (8+3 \pi -6 \ln2)+15 (\pi -\ln4)+10 \alpha (-8+\pi +\ln4)\right)\nonumber\\&+&
\int^1_{0}\frac{Q^2\mu_c z\ln z}{\langle\mathcal{O}_0^{1}\rangle (z^4-1)}dz.
\eea
From (\ref{mu}), we obtain
\be
\mu_{1}-\mu_c\simeq 0.441 \langle\mathcal{O}_0^{1}\rangle^2+0.429 Q^2,
\ee
where we have used the value $\alpha=0.230$ and $\mu_c=1.39$.
Finally, we find
\be
\langle\mathcal{O}_0^{1}\rangle\approx 1.51\sqrt{\mu_{1}-\mu_c-0.429Q^2}.
\ee
This result qualitatively agrees with the numerical curves in Fig.2 and Fig.4. We can see that as the wave number $Q$ increases, the effective critical chemical potential $\tilde{\mu}_c=\mu_c+0.429Q^2$ increases as well, reflecting the fact
that condensate formation becomes harder.
From (\ref{41}), we find that the charge density $\rho_{1}$ may be written as
\be
\rho_{1}=-\frac{1}{2}\langle\mathcal{O}_0^{1}\rangle\chi''(0).
\ee
From the equation of motion (\ref{chi}), we may deduce
\be
\chi''(0)=\frac{1+3z^4}{z-z^5}\chi'(z)\bigg|_{z\rightarrow 0}+\frac{Q^2\mu_c}{\langle\mathcal{O}_0^{1}\rangle}=\frac{1}{z}\chi'(z)\bigg|_{z\rightarrow 0}+\frac{Q^2\mu_c}{\langle\mathcal{O}_0^{1}\rangle}.
\ee
Note that
\be
\frac{1}{z}\chi'(z)\bigg|_{z\rightarrow 0}=-2 \langle\mathcal{O}_0^{1}\rangle \mu_c \int^1_0dz(1-\alpha z^2)^2-\int^1_{1/1000000} \frac{Q^2\mu_c}{\langle\mathcal{O}_0^{1}\rangle z}dz,
\ee
where the lower limit of the second integration is taken to match the numerical computation in the previous subsection.
We find that
\be\label{rho2}
\rho_{1} \approx 2.72(\mu_{1}-\mu_c)+7.74Q^2.
\ee
The above result indicates that even at the critical point
$\mu_{1}=\mu_c$, the charge density $\rho_{1}$ is non-vanishing and there
should be a discontinuity, qualitatively  matching  the
numerical plots presented in Fig.3. When $Q$ is very small, say
$Q=0.01$, it can be easily seen that $\rho_{1}\sim \mathcal{O}(10^{-4})$, which
is negligible. Thus the analytical calculation provides an explanation for the discontinuity in Fig. 3.
Moreover in the $Q\rightarrow 0$ limit, we also recover the result given in \cite{cai1}.

The discontinuity of the charge density may be understood from the condensed matter physics side~\cite{gruner}. According to Peierls theory, we may define a complex order parameter
\begin{equation}
\label{3eq67}
|\Delta|e^{i\varphi}=g(2k_{F})\langle b_{2k_{F}}+b^{+}_{-2k_{F}}\rangle,
\end{equation}
where $b^{+}_{q}, b_{q}$ denote the phonon creation and annihilation operators. The spatially dependent electron density at $T=0$ is given by
\begin{equation}
\rho(x)=\rho_{0}+\frac{1}{\pi}\frac{d\varphi}{dx},
\end{equation}
where $\rho_{0}$ is the electron density in the absence of electron-phonon interaction and $\varphi=\varphi(x,t)$ is the phase of the complex order parameter defined in~(\ref{3eq67}). Hence we see that even as the homogenous charge density $\rho_{0}$
vanishes, the total charge density may still receive corrections from $\varphi(x,t)$. Moreover, as observed for instance from Figure 1 of~\cite{gruner}, distortion may cause a gap in $\epsilon(k)$ at the Fermi level and discontinuity in the charge density. Conversely, the discontinuity found here in the holographic approach also implies that there may be a gap
at the Fermi level.

{$\bullet$ \textsl{Operator dimension $\Delta=\frac{5}{2}$}.}\\
Following the same calculation procedure, we also find that for operator of dimension $\Delta=\frac{5}{2}$ and $\psi_0\sim \langle\mathcal{O}^{2}_0\rangle z^{5/2}F(z) $, the Sturm-Liouville method gives $\mu_{min}=2.67$. The critical value $\mu_c=\mu_{min}\approx 2.67$ is in good agreement with the numerical
value $\mu_c\approx 2.68$ in Figure \ref{condensate8}. However, in the $Q=1$ case the value of $\mu_{c}$ is in less good agreement with the numerical result $\mu_c\approx3.25$ in Figure 2. This discrepancy may be due to the fact that the analytic result does not depend on $Q$ explicitly, while the numerical calculation does depend on $Q$. Therefore good agreement may appear only in the small $Q$ case.

Using the matching method, we obtain
\be
\langle\mathcal{O}^{2}_0\rangle\approx 1.51\sqrt{\mu_{1}-\mu_c-0.82Q^2}.
\ee

The charge density is then given by
\be
\rho\approx 1.322 (\mu_{1}-\mu_c)+16.04 Q^2,
\ee
where $\alpha=0.330$. The above results are qualitatively consistent with the numerical lines in Fig. 3 (right). When $\mu_{1}=\mu_c$, the charge density $\rho_{1}=16.04Q^2$, which also means that there is a jump in the $\rho_{1}-\mu_{1}$ diagram due to nonvanishing $Q$.

\section{Mixing homogeneous and inhomogeneous modes: $A_0(z)\neq 0$ }
We have seen interesting behavior of the charge density even in a
simple case, that is, there exists discontinuity in the
$\rho_{1}-\mu_{1}$ curve for the pure inhomogeneous electrostatic
potential. Let us now turn to the more involved mixed case, including
both the homogeneous and inhomogeneous modes in the electrostatic
potential such that $\mu_0 \neq 0$. In general, we may
 expand both $A_t$ and $\Psi$ in Fourier modes as
\bea
\label{4eq69}
\Psi(x,z)=\sum^{\infty}_{n=0}\psi_n(z)\cos(nQx),\\
A_t(x,z)=\sum^{\infty}_{n=0}A_n(z)\cos(nQx),
\eea
and substitute the above expressions back into (\ref{main1}) and (\ref{main2}). We then obtain a set of coupled non-linear ordinary differential equations for $\psi_n(z)$ and $A_n(z)$,
\bea
&&A''_0+\bigg(\frac{h'}{h}-\frac{1}{z}\bigg)A'_0-\frac{2(\psi^2_0 A_0+2\psi_0 \psi_1 A_1+\frac{1}{2}\psi^2_1 A_1)}{z^2 h}=0,\label{A0}\\
&&A''_1+\bigg(\frac{h'}{h}-\frac{1}{z}\bigg)A'_1-\frac{2(\psi^2_0 A_1+\psi_0 \psi_1 A_0+\frac{3}{4}A_1 \psi^2_1)}{z^2 h}-\frac{Q^2}{h}A_1=0,\label{A11}\\
&&\psi''_0+(\frac{h'}{h}-\frac{3}{z})\psi'_0-\frac{m^2 \psi_0}{z^2 h}
+\frac{1}{h}\bigg((A^2_0+\frac{1}{2}A_1^2)\psi_0+A_0 A_1 \psi_1+\frac{A^2_1}{4}\psi_2\bigg)=0,\label{psi0}\\
&&\psi''_1+(\frac{h'}{h}-\frac{3}{z})\psi'_1-\frac{m^2 \psi_1}{z^2 h}-\frac{Q^2 \psi_1}{ h}\label{psi1}\\
& &+\frac{1}{h}\bigg((A^2_0+\frac{1}{2}A_1^2) \psi_1+2A_0 A_1 \psi_0+A_0A_1\psi_2+\frac{A^2_1}{4}(\psi_1+\psi_3)\bigg)=0,\nonumber\\
&&~~~~~~~~~~~~~~~~~ \vdots\nonumber\\
& &\psi''_n+(\frac{h'}{h}-\frac{3}{z})\psi'_n-\frac{m^2 \psi_n}{z^2 h}-\frac{Q^2 n^2 \psi_n}{h}\\
& &+\frac{1}{h}\bigg[(A^2_0+\frac{1}{2}A^2_1)\psi_n+A_0 A_1(\psi_{(n-1)}+\psi_{(n+1)})+
\frac{1}{4}A^2_1(\psi_{(n-2)}+\psi_{(n+2)})\bigg]=0.\nonumber
\label{psin}
\eea
In the following calculations we will set $A_n(z)=0$
for $1<n\leq n_{max}$. As interpreted in Section 2, here we neglect the higher Fourier modes.

We integrate the equations of motion
(\ref{A0}-\ref{psi1}) from the tip to the boundary. The boundary conditions at the tip are again of Neumann type as given in (\ref{neu1}) and can be expanded as a series of
regular solutions near the tip. We also impose the boundary condition $A_0(0) \rightarrow \mu_{0}-\rho_{0} z^2$, $A_1(0)\rightarrow \mu_{1}-\rho_{1} z^2$ and $A_n(0)=0$ ($1<n<n_{max}$).

\begin{figure}[htbp]
 \begin{minipage}{1\hsize}
\begin{center}
\includegraphics*[scale=0.45] {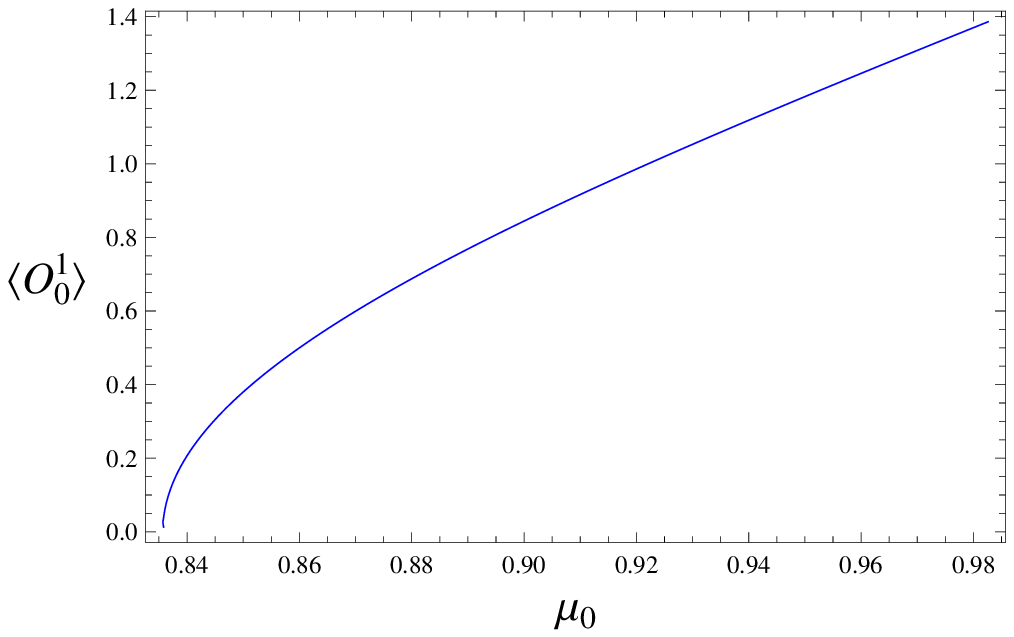}
\includegraphics*[scale=0.46] {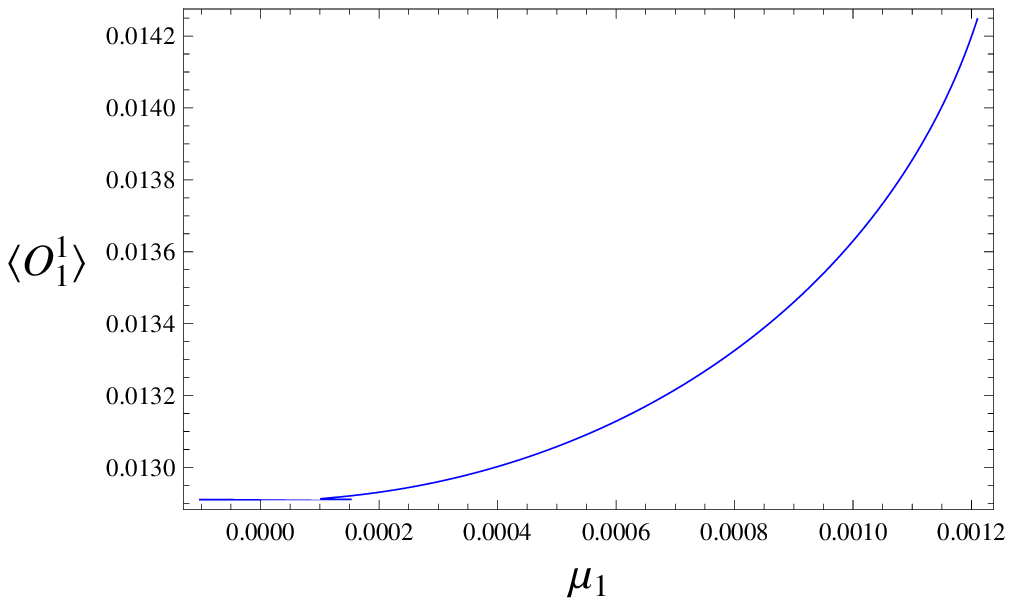}
\includegraphics*[scale=0.44] {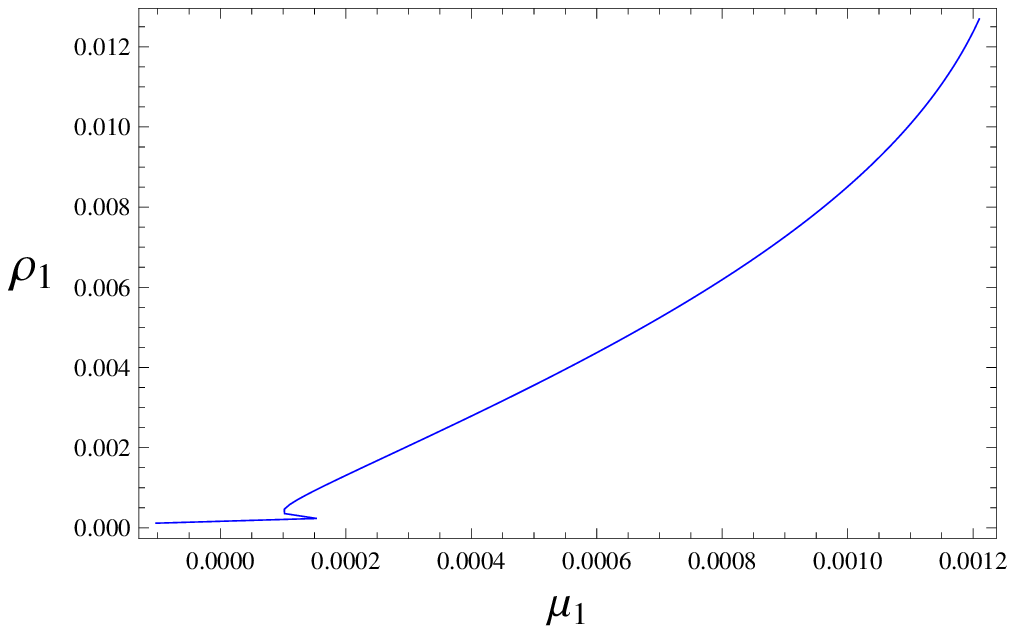}
\end{center}
\caption{(color online) The condensate for the operators $\langle\mathcal{O}^1_0\rangle$ and  $\langle\mathcal{O}^1_1\rangle$ and the charge density $\rho_{1}$ as a function of the chemical potential $\mu_{0}$ and $\mu_{1}$ defined in (2.28), respectively ($Q=0.01$).} \label{conductivity1}
\end{minipage}
\end{figure}

$\bullet ~~~\textsl{Q=0.01}~case.$~~~We plot the condensate for the operators $\langle\mathcal{O}^1_0\rangle$ and $\langle\mathcal{O}^1_1\rangle$ as a function of $\mu_{1}$ in Figure \ref{conductivity1}, keeping in mind that
\be
\langle\mathcal{O}^i\rangle=\langle\mathcal{O}^i_0\rangle+\langle\mathcal{O}^i_1\rangle \cos Qx+...
\ee
We first draw the homogenous condensate $\langle\mathcal{O}^1_0\rangle$ as a function of the chemical potential $\mu_{0}$ in Figure \ref{conductivity1}. For small wave number $Q$, the critical chemical potential is about $\mu_c=0.84$. As compared to the result given in \cite{takayanaki}, a small  $Q$ contributes only minor modifications to the phase diagram.
We also note that for small $Q$, the inhomogeneous operator $\langle\mathcal{O}^1_1\rangle$ is very small compared to $\langle\mathcal{O}^1_0\rangle$. The same is true for the charge density $\rho_{1}$.
The analytic calculation in the next subsection will confirm this. Figure \ref{conductivity1} also implies that the inhomogeneous corrections to the condensate and the charge density are both positive. Consequently, the phase transition is impeded by the presence of spatial modulation.

\begin{figure}[htbp]
 \begin{minipage}{1\hsize}
\begin{center}
\includegraphics*[scale=0.5] {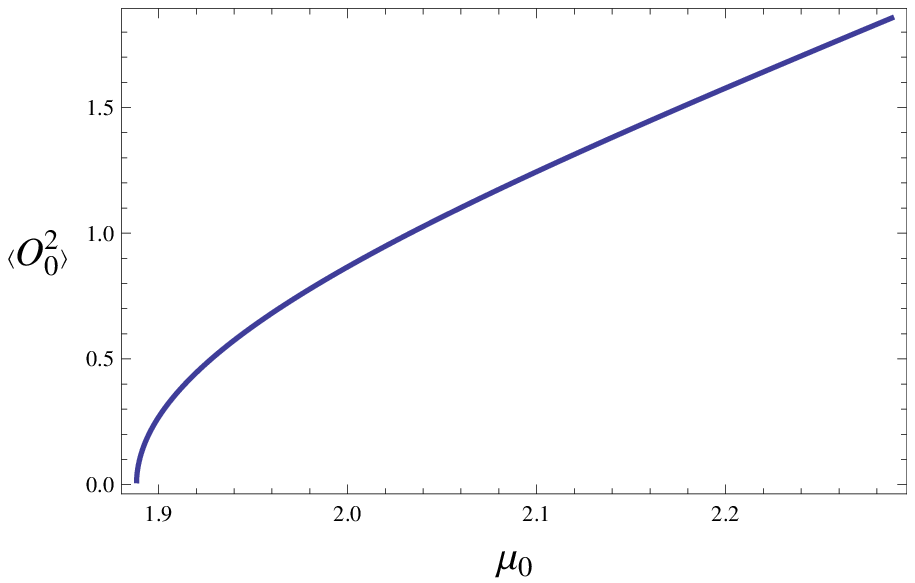}
\includegraphics*[scale=0.5] {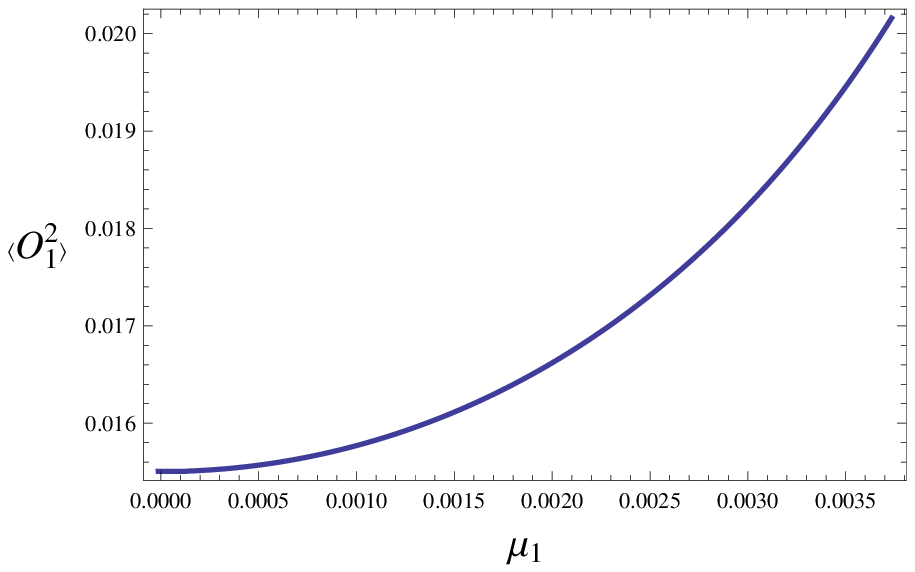}
\includegraphics*[scale=0.5] {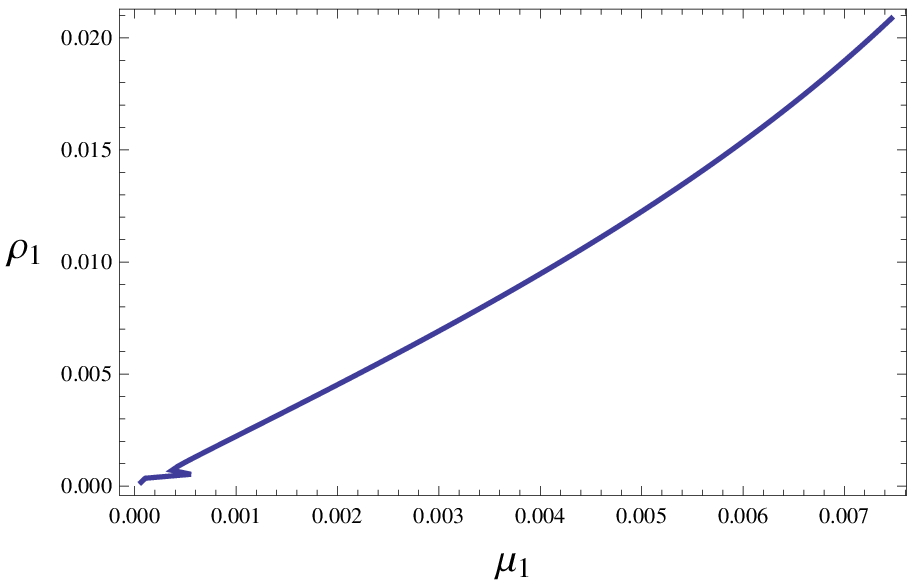}
\end{center}
\caption{(color online) The condensate for the operator $\langle\mathcal{O}^2_0\rangle$ as a function of the chemical potential $\mu_{0}$ (left). The condensate for the operator $\langle\mathcal{O}^2_1\rangle$ (middle) and the charge density $\rho_{1}$ as a function of the chemical potential $\mu_{1}$ (right) ($Q=0.01$).} \label{conductivity2}
\end{minipage}
\end{figure}

\begin{figure}[htbp]
 \begin{minipage}{1\hsize}
\begin{center}
\includegraphics*[scale=0.5] {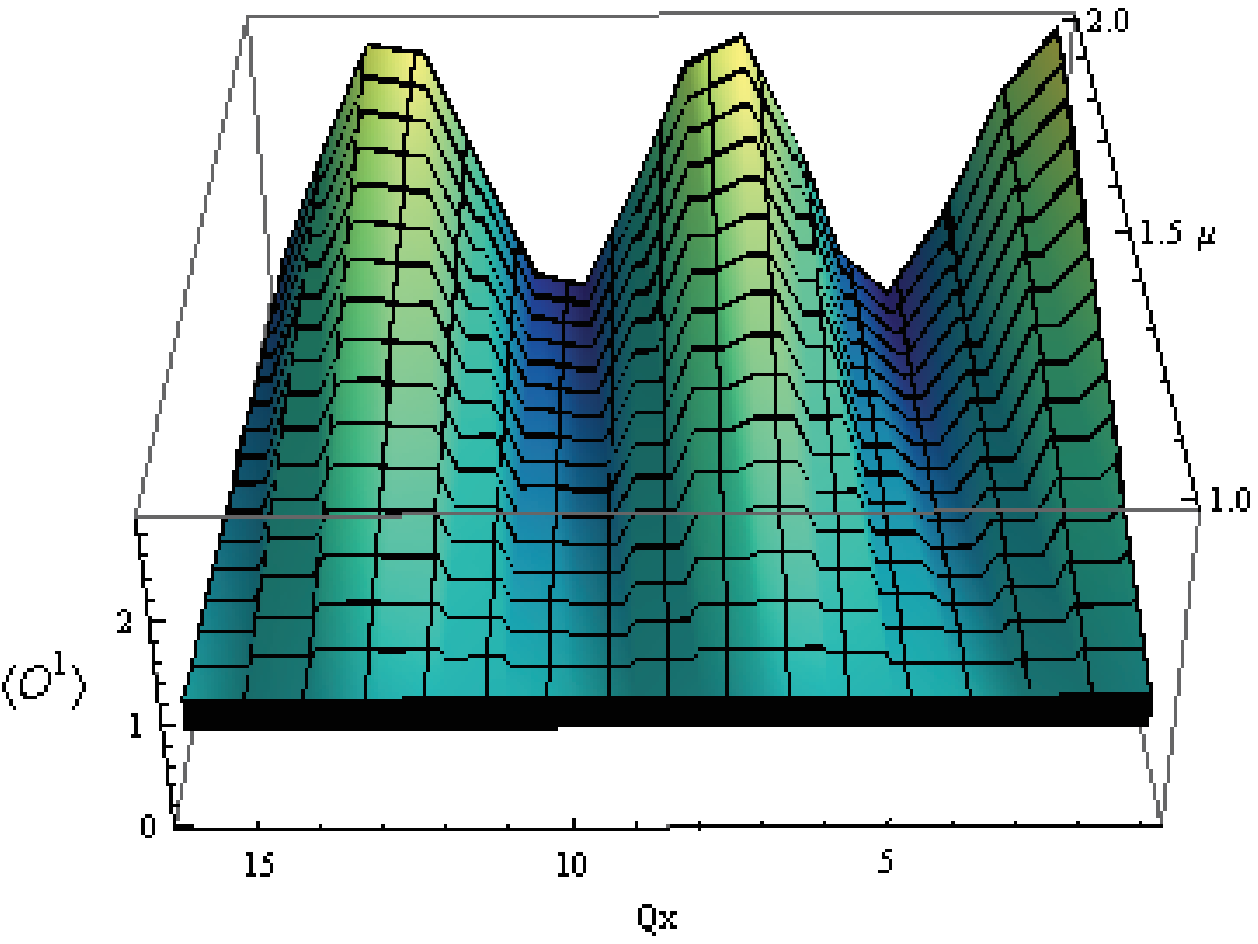}
\includegraphics*[scale=0.5] {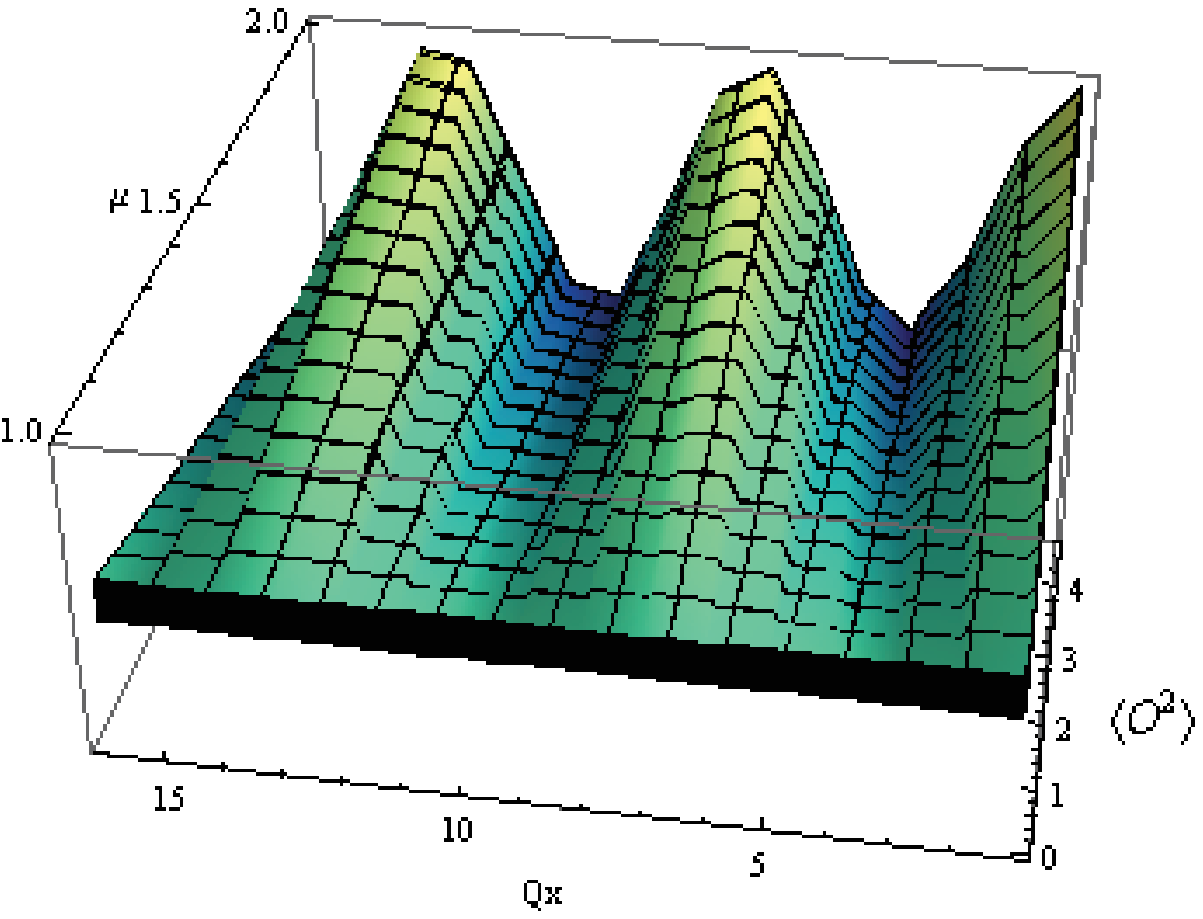}
\end{center}
\caption{(color online)~Q=0.01: The dependence of the superconducting condensate $\langle\mathcal{O}^1 \rangle$ (left) and $\langle\mathcal{O}^2\rangle$  on the chemical potential $\mu=\mu_0+\mu_1\cos Qx$ and $Q x$, where $\langle O^{i}\rangle$ are defined in~(\ref{2eq33}). Note that we are working in units $q=l=1$, and the condensates are given in units of $ql$.} \label{3D1}
\end{minipage}
\end{figure}

\begin{figure}[htbp]
 \begin{minipage}{1\hsize}
\begin{center}
\includegraphics*[scale=0.45] {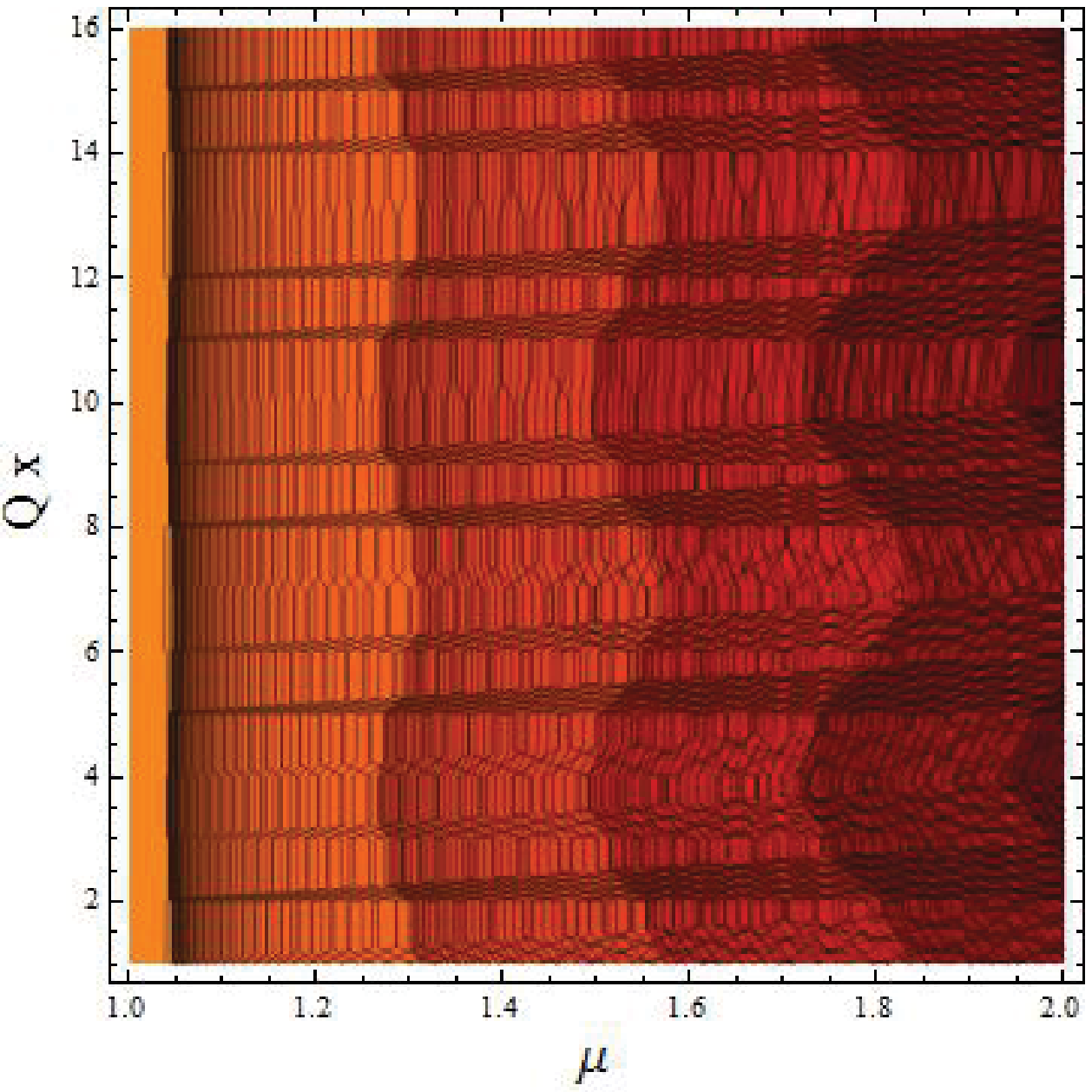}
\includegraphics*[scale=0.45] {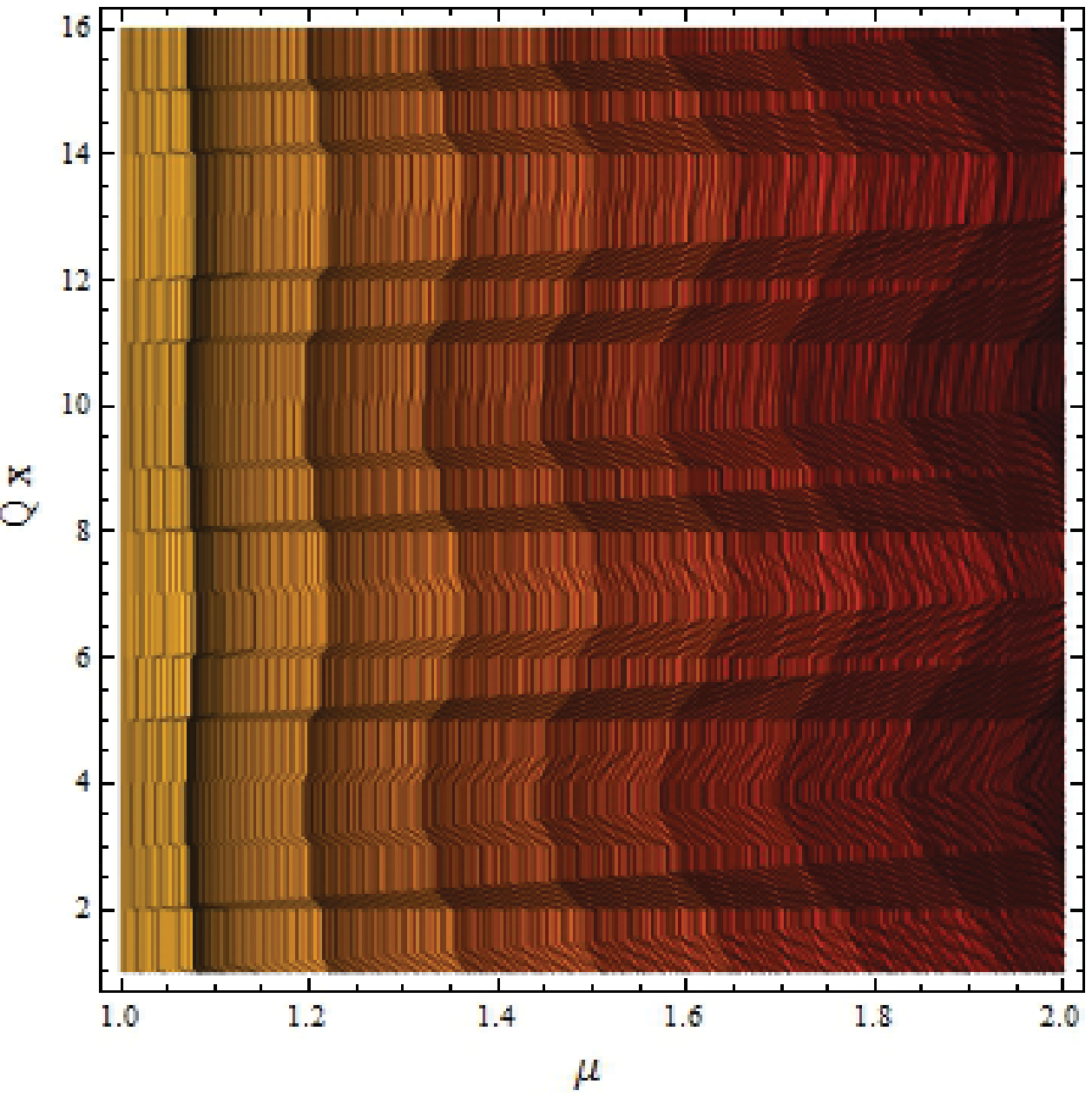}
\end{center}
\caption{(color online)~Q=0.01: The contour plot of the superconducting condensate for  $\langle\mathcal{O}^1\rangle$ (left) and $\langle\mathcal{O}^2 \rangle$ defined in (2.33)~(right). The darker color corresponds to smaller value of condensate in both plots and the stripes correspond to the horizontal lines. } \label{3D}
\end{minipage}
\end{figure}

 We also plot the condensate for the operators $\langle\mathcal{O}^2_1\rangle$ and the charge density $\rho_{1}$ as a function of the chemical potential $\mu_{1}$ ($Q=0.01$) in Figure \ref{conductivity2}. Figure \ref{3D1} shows
  $3$-dimensional plots of our numerical results, while contour plots are shown in figure \ref{3D}. The stripes are clearly visible. Note that for anti-face stripes as seen in realistic cuprates, the electrons for two neighbouring stripes have anti-parallel spins~\cite{emery}. However, here we are discussing the CDW rather than the SDW, we cannot distinguish the spin of the elelctrons. Therefore the stripe diagrams are not the same as the usual anti-face stripe diagram in
  the realistic cuprate superconductors. As seen from Figure 7 (right panel), there appears to be a first order phase transition in the $\rho_{1}-\mu_{1}$ diagram. We confirm the presence of a first order transition below by an analytic calculation. Recall that in the purely inhomogeneous case, the $\rho_{1}-\mu_{1}$ diagram exhibits
  a discontinuity, hence we may argue that the homogeneous part of the electrostatic potential has the effect to remove this discontinuity from the  phase diagram.


$\bullet ~~~\textsl{Q=1}~case. $ As the wave number $Q$ increases, the condensate shows the following interesting behavior.
We shall first consider the $n=0$ terms in the expansions~(\ref{4eq69}). Comparing Figure 6 (left) and Figure 10 (left),  the critical chemical potential in the former case is around 0.837 for $Q=0.01$, while in the latter case, the critical chemical potential  is 1.41 for $Q=1$. Thus larger wave number
may impede the phase transition even in the homogeneous part.

For the $n=1$ terms, the operator $\langle\mathcal{O}^1_1\rangle$ increases as the chemical potential increases. However, the charge density
$\rho_{1}$ decreases as the chemical potential $\mu_{1}$ increases.  We see that there is a jump in the curve of the charge density $\rho_{1}$, which is caused by the inhomogeneity.
Interestingly, the chemical potential contribution $\mu_{1}$, which comes from the inhomogeneous part $A_{1}$, is negative. Still we have $\mu_{0}>|\mu_{1}|$ and there is no discontinuity in the charge density. This negative value may be caused by the interactions between the zeroth order $\mu_{0}$ and first order $\mu_{1}$. The same phenomena are found for the $\Delta=\frac{5}{2}$ case. In contrast, such a negative value is not observed for $A_0(z)=0$ case.
\begin{figure}[htbp]
 \begin{minipage}{1\hsize}
\begin{center}
\includegraphics*[scale=0.6] {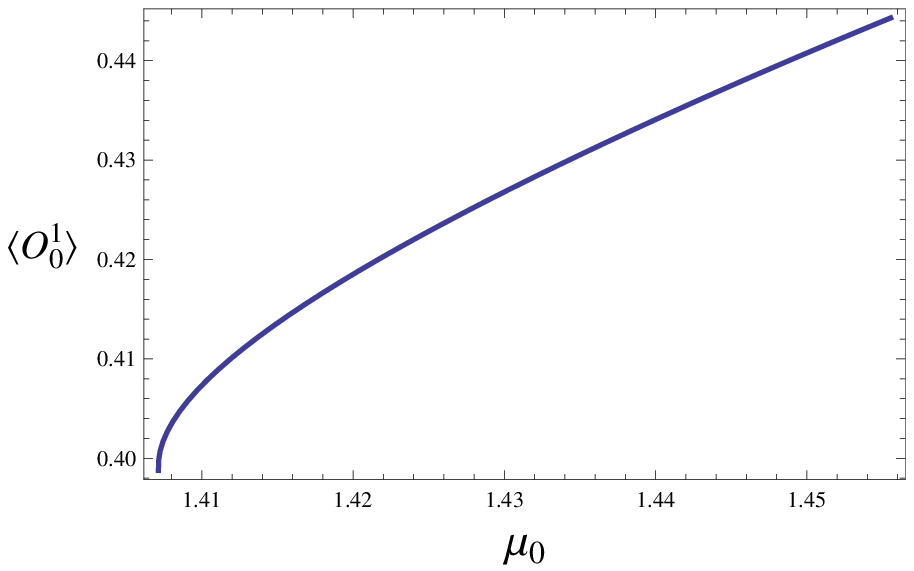}
\includegraphics*[scale=0.58]{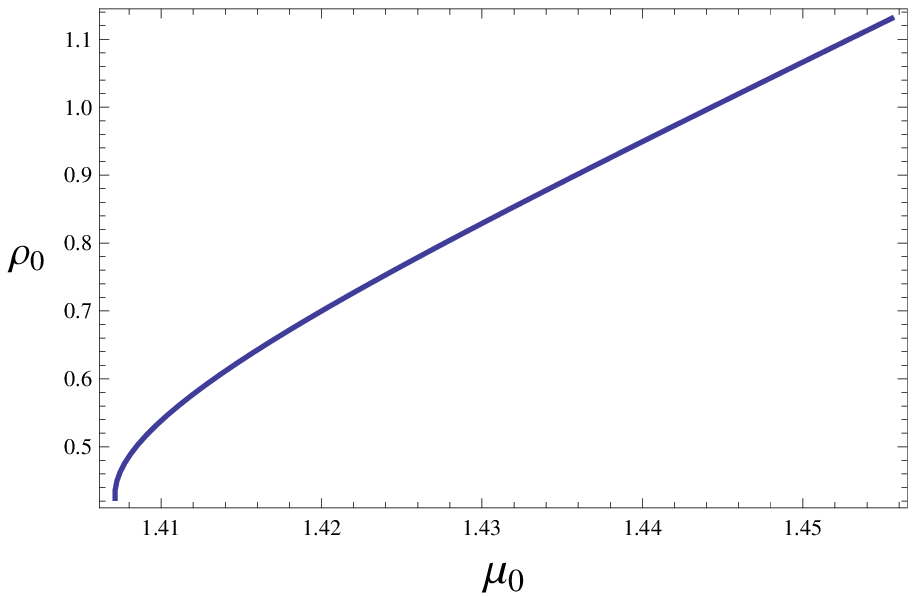}
\includegraphics*[scale=0.60]{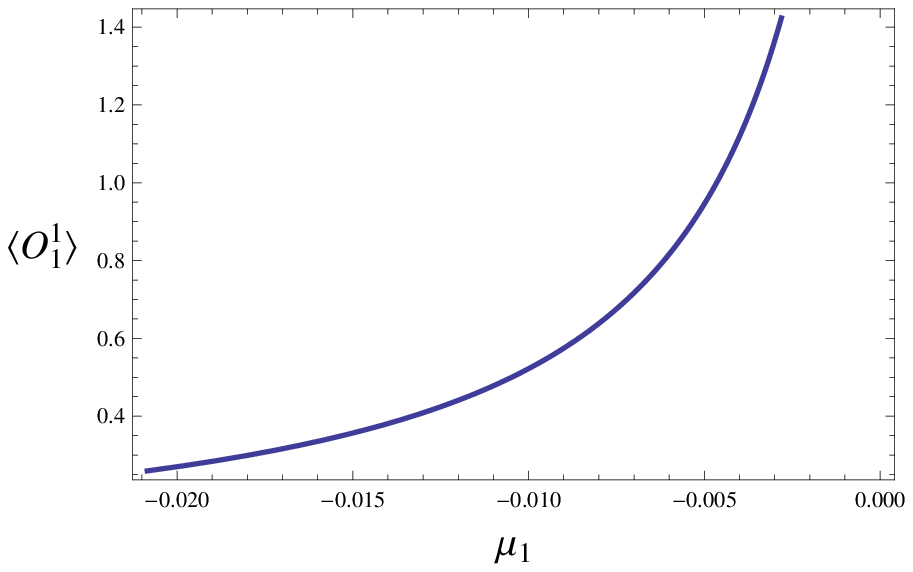}
\includegraphics*[scale=0.6]{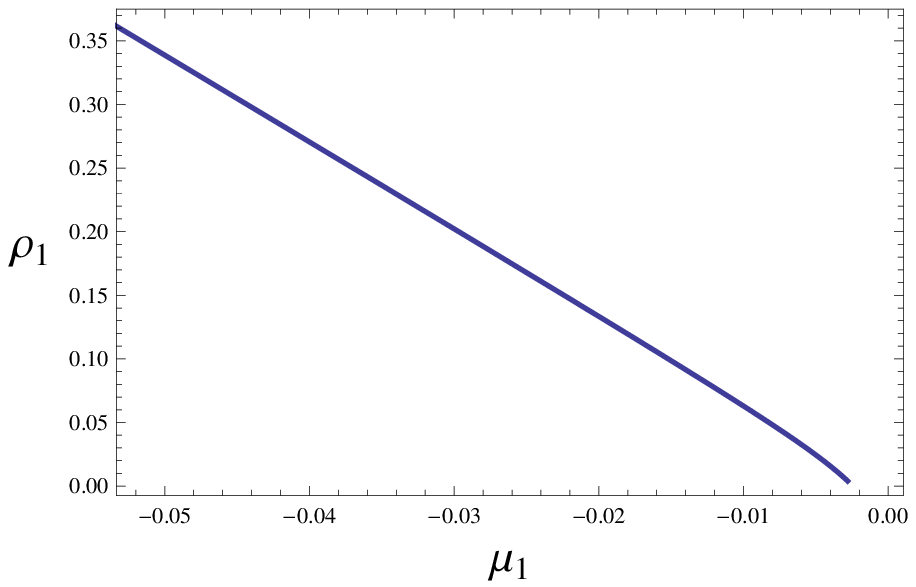}
\end{center}
\caption{(color online) ${Q=1}$ case: The condensate for the operators $\langle\mathcal{O}^1_0\rangle$, $\langle\mathcal{O}^1_1\rangle$ and the charge density $\rho_{1}$ as a function
of the chemical potential $\mu_{0}$ and $\mu_{1}$, respectively.} \label{Q1one}
\end{minipage}
\end{figure}

It is interesting to compare our numerical results to those obtained in~\cite{takayanaki}, where only the homogeneous electrostatic potential is considered. It was found in~\cite{takayanaki} that the critical value of the chemical potential for $\Delta=3/2$ is $\mu_{c}=0.84$, while here $\mu_{0c}\approx0.837$ for small $Q$ and $\mu_{0c}\approx1.40$ for large $Q$ with the same $\Delta$. It seems that for small $Q$, the phase transition receives minor modifications, while for large $Q$ the phase transition requires a much larger $\mu_{0c}$. Therefore our result agrees with the arguments of the introduction, i. e., a large wave vector may suppress the insulator/superconductor transition.
\subsection{Analytical calculation}
As illustrated in previous section, analytical computations can provide insightful additional information to the numerical result.
Here we
analyze the equations of motion to first order by using the Sturm-Liouville eigenvalue method. For simplicity, we only consider conformal dimension
$\Delta=\frac{3}{2}$ case. Let us first assume
\be\label{rel1}
\psi_0\sim z^{3/2}\langle\mathcal{O}_0^{1}\rangle F(z),~~~~\psi_1\sim z^{3/2}\langle\mathcal{O}_1^{1}\rangle F(z),
\ee
so that near the critical point, equations  (\ref{psi0}) and (\ref{psi1}) become
\bea
\bigg[(z^4-1)F'\bigg]'+\frac{9}{4}z^2 F-(\mu^2_0+2\mu_0\mu_{1} \frac{\langle\mathcal{O}_1^{1}\rangle}{\langle\mathcal{O}_0^{1}\rangle}+\frac{\mu^2_1}{2})F=0,\\
\bigg[(z^4-1)F'\bigg]'+\frac{9}{4}z^2 F+Q^2 F-(\mu^2_0+\mu_{0}\mu_{1} \frac{\langle\mathcal{O}_0^{1}\rangle}{\langle\mathcal{O}_1^{1}\rangle}+\frac{3\mu^2_1}{4})F=0.
\eea
Note that $\langle\mathcal{O}_0^{1}\rangle$ and $\langle\mathcal{O}_1^{1}\rangle$ are two undetermined parameters. We shall define the ratio
\begin{equation}
b\equiv\frac{\langle\mathcal{O}_1^{1}\rangle}{\langle\mathcal{O}_0^{1}\rangle},
\end{equation}
which relates the homogeneous condensate $\langle\mathcal{O}_0^{1}\rangle$ and the inhomogeneous condensate $\langle\mathcal{O}_1^{1}\rangle$.  The trial function $F(z)=1-\alpha z^2$ is used to estimate the minimum value.  Then the Sturm-Liouville eigenvalue method gives the minimum values as follows
\bea
&&\bigg(\mu^2_0+2 \mu_{0}\mu_{1} b +\frac{\mu^2_1}{2}\bigg)_{\rm min}=\frac{3}{20}\frac{863\sqrt{230}-14950}{\sqrt{230}-414},\label{p1}\\
&&\bigg(\mu^2_0+\frac{\mu_{0}\mu_{1}}{b}+\frac{3\mu^2_1}{4}\bigg)_{\rm min}=\frac{3}{20}\frac{863\sqrt{230}-14950}{\sqrt{230}-414}+Q^2,\label{p2}
\eea
with  $\alpha\simeq 0.230$ independent of $Q$. We notice that the right hand of (\ref{p1}) takes exactly the same expression as the homogeneous case discussed in \cite{cai1}.
Solving these two equations, we obtain
\bea
&&\mu_{0}=\bigg[24b^4(7+10Q^2)-b^2(133+100Q^2)+28+4K_0\bigg]^{\frac{1}{2}}[10(8-39b^2+48b^4)]^{-\frac{1}{2}},~~~~~~~~~~\\&&\mu_{1}=2[5b^2(7+Q^2)-7-6b^4(7+Q^2)-K_0]\mu_{0}[b(3b^2-1)(20Q^2-7)]^{-1},
\eea
where $K_0=\bigg[\left(1-3 b^2\right)^2 \left(49+4 b^4 \left(7+10 Q^2\right)^2-2 b^2 \left(98+105 Q^2+100 Q^4\right)\right)\bigg]^{1/2}$. As $Q \rightarrow 0$, we recover the result obtained in \cite{cai1} with $\mu_{0}\simeq 0.837$
and $\mu_{1}=0$. Because of the mixing of the Fourier modes, the zeroth order and the first order condensate may
interact with each other. This can be seen clearly from Table \ref{table1} that for larger ratio $b$ and larger wave number $Q$, the critical chemical potential $\mu_{0}$ increases, but $\mu_{1}$ drops to  even  a negative-valued number.
This result agrees with the previous numerical results that $\mu_{1}$  can take a negative value. Note that
for $b=0.5$  and $Q=1$, there are no real number solutions to the equations (\ref{p1}-\ref{p2}). The ratio $b$ and $Q$ form a parameter space and the situation we are studying is much more complicated than the homogeneous case.
In the following, we mainly consider the case $b=1$ with different wave numbers $Q=0.001$ and  $Q=1$.
\\
 \begin{table}[ht]
 \centering
 \scriptsize
\caption{\label{CriticalZheng1} The critical chemical potential
$\mu_{0}$ (left column) and $\mu_{1}$ (right column) obtained by the analytical Sturm-Liouville eigenvalue method.}
\begin{tabular}{c c c c c }
         \hline
$b$ & 0.01 & 0.1 & 0.5 & 1
        \\
        \hline
$Q=0.01$~~&$0.837$~~~$1.2\times 10^{-6}$~~~&$0.837$~~~~$1.2\times 10^{-5}$~~~&
$0.837$~~~$1.2\times10^{-4}$~~&
$0.837$~~~$-1.2\times10^{-4}$
          \\
$Q=0.1$~~&$0.837$~~~$1.2\times 10^{-4}$~~~&$0.837$~~~~$1.2\times 10^{-3}$~~~&
$0.831$~~~$1.2\times10^{-2}$~~&
$0.848$~~~$-1.2\times10^{-2}$
          \\
$Q=0.25$~~&$0.837$~~~$7.5\times 10^{-4}$~~~&$0.836$~~~~$7.6\times 10^{-3}$~~~&
$0.796$~~~$7.7\times10^{-2}$~~&
$0.906$~~~$-6.8\times10^{-2}$
          \\
$Q=0.5$~~&$0.837$~~~$3.0\times 10^{-3}$~~~&$0.833$~~~~$3.0\times 10^{-2}$~~~~~&
$0.649$~~~~$0.341$~~&
$1.073$~~~~~~$-0.222$
          \\
$Q=1$~~&$0.836$~~~$1.2\times 10^{-2}$~~~&$0.820$~~~~$1.2\times 10^{-1}$~~~&
~~~$--$~~~~$--$~~~~&~~
$1.532$~~~~~~$-5.95\times10^{-1}$
          \\
        \hline\label{table1}
\end{tabular}
\end{table}

\emph{$\bullet$ Relations of $\langle\mathcal{O}^i\rangle$ -$(\mu-\mu_i)$ and $\rho_i$-$(\mu-\mu_i)$.}\\
It is interesting to investigate the relation  of $\langle\mathcal{O}^i\rangle$ -$(\mu-\mu_i)$ and $\rho_i$-$(\mu-\mu_i)$ analytically and see the mixing of the homogeneous operator $\langle\mathcal{O}^1_0\rangle$ and
inhomogeneous operator  $\langle\mathcal{O}^1_1\rangle$ at different values of the wave number $Q$.

When the chemical potential is slightly above the critical value $\mu_i$, we can rewrite equations (\ref{A0}) and (\ref{A11}) by using the relation, (\ref{rel1})
\bea
&&A''_0+\bigg(\frac{h'}{h}-\frac{1}{z}\bigg)A'_0-\frac{2(c^2_0 A_0+2c_0 c_1 A_1+\frac{1}{2}c^2_1 A_1)zF^2}{ h}=0, \\
&&A''_1+\bigg(\frac{h'}{h}-\frac{1}{z}\bigg)A'_1-\frac{2(c^2_0 A_1+c_0 c_1 A_0+\frac{3}{4}A_1 c^2_1 )zF^2}{ h}-\frac{Q^2}{h}A_1=0,
\eea
where $c_0=\langle\mathcal{O}^1_0\rangle$ and $c_1=\langle\mathcal{O}^1_1\rangle$.
Near the critical value, we expand the potentials $A_0$ and $A_1$ in a power series of the condensate as
\bea
A_0\sim \mu_{0c}+c_0\chi_0(z)+...,~~~A_1\sim \mu_{1c}+c_1\chi_1(z)+...
\eea
Both of the functions $\chi_i(z)$ obey the  boundary condition at the tip $\chi_i(1)=0$. The equations of motion for $\chi_i(z)$  can be expressed as
\bea
&&\chi''_0-\frac{1+3z^4}{z-z^5}\chi'_0=\frac{2(c_0\mu_{0c}+2c_1\mu_{1c}+\frac{c^2_1}{2c_0}\mu_{1c})zF^2}{h},\\
&&\chi''_1-\frac{1+3z^4}{z-z^5}\chi'_1=\frac{2(c_0\mu_{1c}+c_1\mu_{0c}+\frac{3c^2_1}{4c_0}\mu_{1c})zF^2}{h}+\frac{Q^2\mu_{1c}}{c_1 h}.
\eea
Following the procedure given in section 3.2, we evaluate the condensate at $Q=0.01$ to be
\bea
&&\langle\mathcal{O}_0^{1}\rangle=1.940\sqrt{\mu_0-\mu_{0c}+1.900\times10^{-5}\langle\mathcal{O}_1^{1}\rangle^2}+1.428\times 10^{-4}\langle\mathcal{O}_1^{1}\rangle,~~~\\
&&\langle\mathcal{O}_1^{1}\rangle\simeq 19.716 \sqrt{\mu_1-\mu_{1c}},
\eea
where the corresponding critical chemical potentials are $\mu_{0c}=0.837$ and $\mu_{1c}=-0.0001$.
On the other hand, the condensate at $Q=1$ is given by
\bea
&&\langle\mathcal{O}_0^{1}\rangle=1.434\sqrt{\mu_0-\mu_{0c}+0.168\langle\mathcal{O}_1^{1}\rangle}+0.389\langle\mathcal{O}_1^{1}\rangle,\\
&&\langle\mathcal{O}_1^{1}\rangle \simeq 3.923\sqrt{\mu_1- \mu_{1c}},
\eea
where $\mu_{0c}=1.531$ and $\mu_{1c}=-0.595$.
The charge density $\rho_i$ can be evaluated by using the ansatz $\rho_i=-\frac{1}{2}\langle\mathcal{O}_i^{1}\rangle\chi''_i(0)$, i. e.
\bea
\rho_{0}=2.700(\mu_0-\mu_{0c})+3.86(\mu_0-\mu_{0c})Q^2,\\
\rho_{1}=278.870(\mu_1-\mu_{1c})-0.004\sqrt{\mu_0-\mu_{0c}}\sqrt{\mu_1-\mu_{1c}}+0.00077 Q^2,
\eea
where $\mu_{0}=0.837$ and $\mu_{1}=-0.0001$.
The analytical computation may not be able to match the numerical result exactly, but it can qualitatively explain the behavior of the condensate and the charge density. In particular, it can easily be seen that
$$
\frac{\partial\rho_{1}}{\partial\mu_{1}}\sim\frac{1}{\sqrt{\mu_{1}-\mu_{1c}}},
$$
which means that the first order derivative of $\rho_{1}$ with respect to $\mu_{1}$ diverges as $\mu_{1}\rightarrow\mu_{1c}$. Therefore the first order phase transition in Figure 6 and 7 may be understood in the context of the analytical results presented here.
\section{The Grand  Canonical Potential}
Here we study the grand potential $\Omega$ in the grand canonical ensemble, which is the Legendre transform of the free energy $F$. Following~\cite{Flauger:2010tv}, we are interested in the action for $A_t$ and $\Psi$ only. The action for the matter fields is given by
\be
S_M=\int d^5x\sqrt{-g}\bigg(-\frac{1}{4}F^{\mu\nu}F_{\mu\nu}-|\partial_{\mu}\Psi-iqA_{\mu}\Psi|^2
-m^2|\Psi|^2\bigg).
\ee
In the probe limit, the ADM energy density is not included in the action.
We integrate $S$ by parts and use the equations of motion to evaluate the on-shell action
\bea
&&S_M=\int d^5 x [e.o.m.]+S_{on-shell},\\
&&S_{on-shell}=-\int d^4 x \frac{h(z)}{2 z}A_t A'_t+\int_{z=0} d^4 x \frac{h(z)}{z^3}\Psi \Psi'-\int d^5 x \frac{A^2_t \Psi^2}{z^3},\label{onshell}
\eea
where we have used $h(1)=0$ for $h(z)$ as in (2.20). Recall that the boundary behavior of $\Psi$ and $A_{t}$ is given by $\Psi(z\rightarrow 0)=\Psi^{(1)}{z^{3/2}}+\Psi^{(2)}{z^{5/2}}$ and
$A_t(z\rightarrow 0)=\mu-\rho z^2$. We find that only the second term in (\ref{onshell}) is divergent, which can be regularized by introducing a cutoff $z=\epsilon$
\be
S_{on-shell}=\int d^4x \frac{3}{2}\frac{\Psi^{(1)}}{\epsilon}+\rm{finite~~ terms},
\ee
and adding the following boundary counter term
\be
S_{ct}=-\int d^3 x\frac{h}{z^3}\Psi^2.
\ee
The renormalized grand canonical potential is then
\be
\Omega=-S_M=\int d^3 x\bigg[\frac{h}{2 z}A_t A'_t-\frac{h}{z^3}\Psi\Psi'
+\frac{3}{2}\frac{h}{z^4}\Psi^2\bigg]+\int d^4 x \frac{A^2_t \Psi^2}{z^3}.
\ee
\begin{figure}[htbp]
 \begin{minipage}{1\hsize}
\begin{center}
\includegraphics*[scale=0.6] {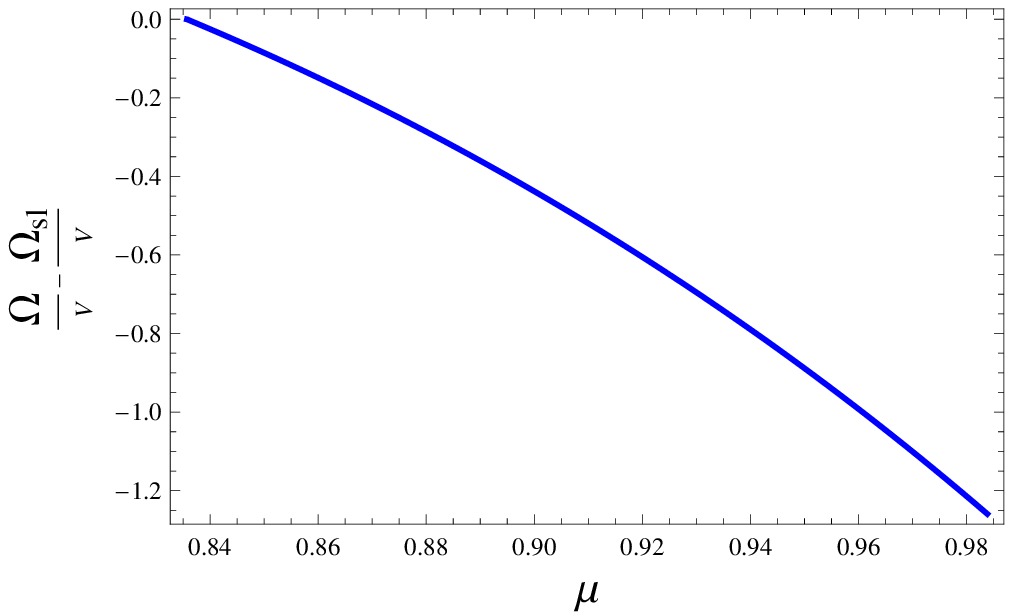}
\includegraphics*[scale=0.6] {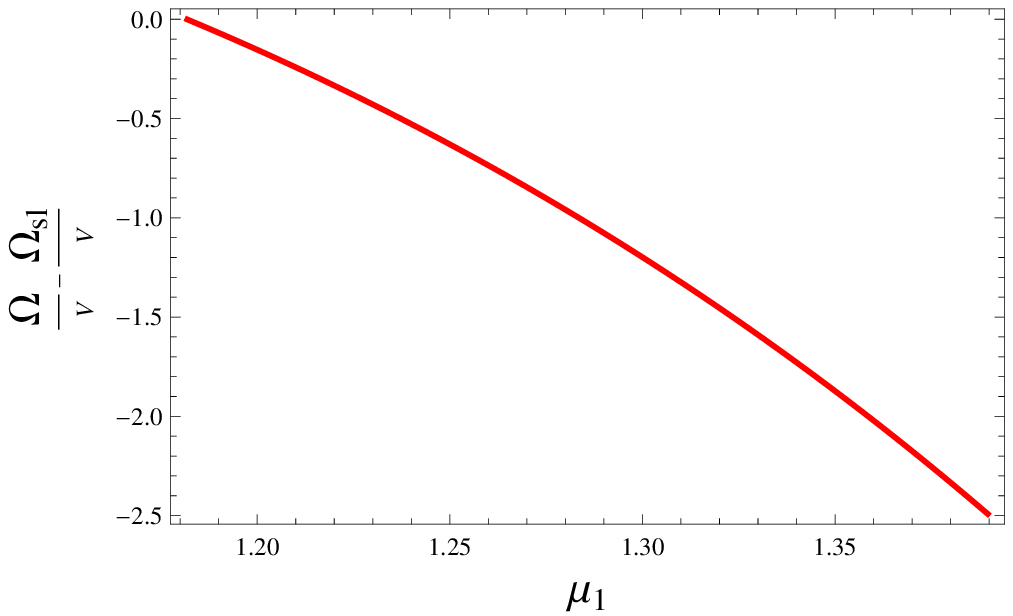}
\end{center}
\caption{(color online)~The grand canonical potential per unit $F=\frac{\Omega}{V}$ against chemical potential. Left:  The grand canonical potential  corresponds to $\langle\mathcal{O}^1\rangle$ for $A_0\neq 0$.
Right: The grand canonical potential corresponds to $\langle\mathcal{O}^1\rangle$ for $A_0=0$. We choose $Q=1/100$ for both cases.} \label{freeenergy}
\end{minipage}
\end{figure}
The average grand canonical potential per unit volume can be expressed as
\be\label{grand}
\frac{\Omega}{V}=\frac{Q}{2\pi}\int_{z=0}\frac{dx}{2 z}A_t A'_t+\frac{Q}{2\pi}\int dz dx\frac{A^2_t \Psi^2}{z^3}.
\ee
We may also evaluate the grand canonical potential $\Omega$ in the insulator phase where $\Psi=0$. It was found in \cite{takayanaki} that for the homogenous AdS soliton without scalar field, the entropy and charge density vanish, so the grand canonical potential is determined  by the ADM energy density
\be
\frac{\Omega_{sl}}{V}=-\frac{\pi l^3}{R^3_0}=-1.
\ee
We may compute (\ref{grand}) for the soliton state profile of $A_t$ and obtain a vanishing grand canonical potential in the probe limit, i.e.  $\frac{\Omega_{sl}}{V}=0$, since for the soliton state both the charge density and the scalar field $\Psi$ vanish.
In figure \ref{freeenergy}, we show that the grand potential $F=\Omega/V$ is lower than the insulating phase ${\Omega_{sl}}/V$ and is therefore favored. The same results can be obtained for the $Q=1$ cases.
\section{Conductivity perpendicular to the direction of the stripes}
For completeness we study the conductivity in the presence of spatially modulated electrostatic potential. At first let us consider the simpler case, that is, the conductivity $\sigma_y$ perpendicular to the direction of the stripes.  Introducing a small perturbation
\be
A_y=\int \frac{d\omega dk}{(2\pi)^3}A_y(z,\omega,k)e^{i(kx-\omega t)},
\ee
the equation of motion for $A_y(z,\omega,k)$ then reads
\be
A''_y+(\frac{h'}{h}-\frac{1}{z})A'_y+\frac{\omega^2}{h}A_y-\frac{k^2}{h}A_y=\frac{2 A_y}{z^2 h}\psi^2.
\ee
Note that we should still impose the Neumann boundary condition at the tip $r=r_{0}$. The asymptotic behavior of $A_y$ near $z\rightarrow 0$ goes as
\be
A_y=A^{(0)}_y(x)+A^{(1)}_y(x)z^2+\frac{A^{(0)}_y(x) \omega^2}{2} z^2 \log \frac{\Lambda}{z}.
\ee
The optical conductivity in the $y$-direction is given by
\be
\sigma_y(\omega,k=0,x)=\frac{-2i A^{(1)}_y(x)}{\omega A^{(0)}_y(x)}+\frac{i\omega}{2}.
\ee

\begin{figure}[htbp]
 \begin{minipage}{1\hsize}
\begin{center}
\includegraphics*[scale=0.86] {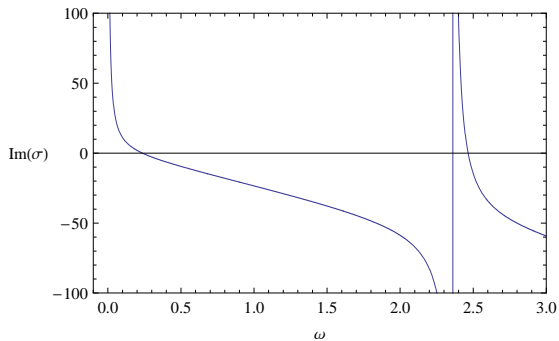}
\end{center}
\caption{(color online)~The imaginary part of the conductivity for the AdS soliton with a scalar condensate with $\rho=0.32$ and $\mu=0.86$.} \label{conductivity}
\end{minipage}
\end{figure}

Our numerical calculations show that the real part of the conductivity vanishes, which means that there is no dissipation and is consistent with the
absence of horizon. The imaginary part of $\sigma_y$ is plotted in Fig.\ref{conductivity}. We observe a pole at $\omega=0$. The Kramers-Kronig relation
\be
Im \sigma_y=\frac{1}{\pi}\int^{0}_{-0} d \omega'\frac{Re \sigma_y}{\omega-\omega'}
\ee
leads to the fact that $Re \sigma_y$ is just a series of  delta functions. This result is consistent with the observation in \cite{takayanaki} that the main difference between the black hole superconductor and soliton superconductor
 is the conductivity: For the black hole superconductor, the low temperature conductivity has a gap at low frequency, but approaches the normal state conductivity at larger frequency. In brief, the conductivity
 perpendicular to the direction of the stripe is the same as in the homogeneous case \cite{takayanaki}.

 The conductivity parallel to the direction of the stripes $\sigma_x$ is much more difficult to evaluate than the conductivity discussed above.  The computation of $\sigma_x$ is complicated because the inhomogeneity is
 in the $x-$direction and applying an electric field in the $x-$direction sources other independent perturbations even at linear order.
 Particularly, the imaginary part of the complex scalar field $\Psi$ will appear in the equations and we have to solve partial differential equations. We will leave this study to future work.

\section{Conclusion}
We have studied the striped phase and the CDW in the holographic insulator/superconductor transition by considering a spatially modulated chemical potential, where both $U(1)$ gauge symmetry and translational symmetry are broken. We first consider the simpler case with a pure inhomogeneous $A_{t}$, where we find that the presence of the inhomogeneity increases the value of the critical chemical potential and makes the phase transition more difficult to occur. Moreover, a discontinuity in the charge density as a function of the chemical potential may also be attributed to the inhomogeneity. We confirm these arguments by analytical methods. Note that a similar discontinuity was observed in the backreacted background in~\cite{horowitz2}, where the reason for the discontinuity was unclear. It would be interesting to see if such a discontinuity still exists if backreaction is taken into account in our setup.

Subsequently we turned to the more complicated case with both homogeneous and inhomogeneous electrostatic potentials included. We find that the discontinuity in the charge density still exists and the contribution to the chemical potential associated with the inhomogeneous electrostatic potential may become negative. The former observation appears also to be due to the inhomogeneity and the latter may be related to the interaction between the homogeneous and the inhomogeneous parts of the gauge potential. These arguments are also confirmed via analytical methods, at least qualitatively. We evaluate the grand canonical potential and find that the striped phase is favored. Finally for completeness we also study the conductivity perpendicular to the direction of the stripe and find precise agreement with the homogeneous case.

The fact that the spatially modulated chemical potential disfavors the phase transition may be understood by studying the effective mass of the charged scalar, which is given by
\be
m^2_{\rm eff}=m^2+g^{xx}Q^2n^2+g^{tt}q^2A^2_t.
\ee
In the homogeneous case, $Q=0$ and the last term is negative. Once the
effective mass becomes negative, there
 is an instability towards developing non-trivial scalar hair. However, the second term is positive in the presence of the inhomogeneity. For fixed charge $q$ and large enough wave vector $Q$, the effective mass may become positive, which prevents the instability from occurring.

To conclude, let us discuss some future directions. First of all, it
would be interesting to realize spontaneously generated striped phases
in the holographic insulator/superconductor transition with
spontaneous breaking of translational invariance, along the lines
of~\cite{Donos:2011bh} and~\cite{Donos:2013gda}. Next, it would be
interesting to investigate SDW in the holographic superconductor
models. Magnons are the collective excitations of the SDW ground state
with well-defined magnetic characters and the SDW  ground state is
closely related to the antiferromagnetic order in the Mott
insulators. However, since in this paper we are dealing with a
spatially modulated source which does not have magnetic character, we
do not expect that the antiferromagnetic order can be described by the
model presented here in a straightforward generalization. A
holographic realization of SDW would allow for an identification of
the AdS insulator/superconductor transition with the Mott
insulator/superconductor transition observed in cuprates. Finally, for
homogeneous cases, it has been pointed out in~\cite{takayanaki} that
even in the probe limit, the phase diagram of AdS soliton and AdS
black hole with a charged scalar field is analogous to the phase
diagrams of the electron-doping cuprate superconductors. A complete
phase diagram was obtained in~\cite{horowitz2} by considering the
backreaction. In this paper, we worked in the probe limit and hence
are not able to explore the complete phase diagram. Therefore a more
ambitious goal is to take the
backreaction into account and investigate the complete
phase diagram of the holographic insulator/superconductor transition in the presence of stripes.

{\bf Note Added}: While finalizing this work, we received the paper~\cite{Domokos:2013kha}, where the authors introduce a novel set of stability conditions for spatially modulated phases. It would be interesting to perform a similar analysis on our spatially modulated soliton background.

\vspace*{10mm} \noindent
 {\large{\bf Acknowledgments}}

\vspace{1mm}  We would like to thank Jerome Gauntlett and Koenraad Schalm for discussions and Jonathan Shock for kind help with the numerical simulations. The work of J.E. was supported in part by the DFG cluster of excellence `Origin and Structure of the Universe'
(www.universe-cluster.de).   The work of XHG
was partly supported by the NSFC (No. 11005072) and
Shanghai Rising-Star Program (10QA14023000). DWP is supported by the Alexander von Humboldt Foundation. XHG would like to thank Max-Planck institute for physics for warm hospitality.

\vspace{1mm}


\renewcommand{\theequation}{A.\arabic{equation}}

\setcounter{equation}{0} \setcounter{footnote}{0}


\begin{thebibliography}{99}
\bibitem{ads/cft}
J. M. Maldacena, {Adv. Theor. Math. Phys.} {\bf 2} (1998) 231, {
[arXiv:hep-th/9711200]}.

\bibitem{gkp}
S. S. Gubser, I.R. Klebanov and A.M. Polyakov, Phys.\ Lett.\ {\bf
B428} (1998) 105, { [arXiv:hep-th/9802109]}.

\bibitem{w}
E. Witten, Adv.\ Theor.\ Math.\ Phys.\ {\bf 2} (1998) 253, {
[arXiv:hep-th/9802150]}.
\bibitem{Hartnoll:2008vx}
  S.~A.~Hartnoll, C.~P.~Herzog and G.~T.~Horowitz,
  ``Building a Holographic Superconductor,''
  Phys.\ Rev.\ Lett.\  {\bf 101}, 031601 (2008)
  [arXiv:0803.3295 [hep-th]].
\bibitem{Hartnoll:2008kx}
  S.~A.~Hartnoll, C.~P.~Herzog and G.~T.~Horowitz,
  ``Holographic Superconductors,''
  JHEP {\bf 0812}, 015 (2008)
  [arXiv:0810.1563 [hep-th]].
\bibitem{gub1}
  S.~S.~Gubser,
  ``Phase transitions near black hole horizons,''
  Class.\ Quant.\ Grav.\  {\bf 22}  (2005) 5121.
  [arXiv:hep-th/0505189].

\bibitem{gub2}
S.~S.~Gubser,
  ``Breaking an Abelian gauge symmetry near a black hole horizon,''
  Phys.\ Rev.\ D {\bf 78}, 065034 (2008)
  [arXiv:0801.2977 [hep-th]].
\bibitem{takayanaki} T. Nishioka, S. Ryu and T. Takayanagi, ``Holographic superconductor/insulator transition at zero temperature,''
JHEP {\bf 1003}  (2010) 131 [arXiv:0911.0962[hep-th]]
\bibitem{horowitz2} G.~T.~Horowitz and B.~Way,
  ``Complete Phase Diagrams for a Holographic Superconductor/Insulator System,''
  JHEP {\bf 1011}, 011 (2010)
  [arXiv:1007.3714 [hep-th]].
\bibitem{martin} I. Martin, D. Podolsky, and S. A. Kivelson,
``Enhancement of superconductivity by local inhomogeneity'',
Phys. Rev. B {\bf 72} 060502 (2005) [arXiv:cond-mat/1110.4632].
\bibitem{li}Q. Li, M. Hucker, G. D. Gu, A. M. Tavelik and J. M. Tranquada, ``Two-Dimensional Superconducting Fluctuations in Stripe-Ordered $La_{1.875}Ba_{0.125}CuO_{4}$'',
Phys. Rev. Lett. 99 (2007) 067001
\bibitem{berg} E. Berg, E. Fradkin, S. A. Kivelson and J. M. Tranquada, ``Striped superconductors: how spin, charge and superconducting orders intertwine in the cuprates'',
New J. Phys. 11 (2009) 115004.
\bibitem{emery} V. J. Emery, S. A. Kivelson and J. M. Tranquada, ``Stripe phases in high-temperature superconductors'', Proc. Natl. Acad. Sci. 96 (1999) 8814
\bibitem{tamada} K. Tamada,
``Doping dependence of the spatially modulated dynamical spin correlations and the superconducting-transition temperature in $La_{2-x}Sr_{x}CuO_4$'', Phys. Rev. B. {\bf 57} (1998) 6165
\bibitem{gruner} G. Gruner, ``The dynamics of charge-density waves'',
Rev. Mod. Phys. 60 (1988) 1129
\bibitem{Donos:2011bh}
  A.~Donos and J.~P.~Gauntlett,
  ``Holographic striped phases,''
  JHEP {\bf 1108}, 140 (2011)
  [arXiv:1106.2004 [hep-th]].
\bibitem{Donos:2011qt}
  A.~Donos, J.~P.~Gauntlett and C.~Pantelidou,
  ``Spatially modulated instabilities of magnetic black branes,''
  JHEP {\bf 1201}, 061 (2012)
  [arXiv:1109.0471 [hep-th]].
\bibitem{Donos:2011ff}
  A.~Donos and J.~P.~Gauntlett,
  ``Holographic helical superconductors,''
  JHEP {\bf 1112}, 091 (2011)
  [arXiv:1109.3866 [hep-th]].
\bibitem{Donos:2012gg}
  A.~Donos and J.~P.~Gauntlett,
  ``Helical superconducting black holes,''
  Phys.\ Rev.\ Lett.\  {\bf 108}, 211601 (2012)
  [arXiv:1203.0533 [hep-th]].
\bibitem{Donos:2012wi}
  A.~Donos and J.~P.~Gauntlett,
  ``Black holes dual to helical current phases,''
  Phys.\ Rev.\ D {\bf 86}, 064010 (2012)
  [arXiv:1204.1734 [hep-th]].
\bibitem{h4} M. Rozali, D. Smyth, E. Sorkin, and J. B. Stang, ``Holographic Stripes,"
arXiv:1211.5600 [hep-th].
\bibitem{h8} N. Iizuka and K. Maeda, ``Stripe Instabilities of Geometries with Hyperscaling
Violation," arXiv:1301.5677 [hep-th].
\bibitem{Donos:2013gda}
    A.~Donos and J.~P.~Gauntlett,
    ``Holographic charge density waves,''
    arXiv:1303.4398 [hep-th].
\bibitem{h5} A. Donos, ``Striped phases from holography," arXiv:1303.7211 [hep-th].
\bibitem{Withers:2013loa}
  B.~Withers,
  ``Black branes dual to striped phases,''
  arXiv:1304.0129 [hep-th].
\bibitem{Withers:2013kva}
  B.~Withers,
  ``The moduli space of striped black branes,''
  arXiv:1304.2011 [hep-th].

\bibitem{Flauger:2010tv}
  R.~Flauger, E.~Pajer and S.~Papanikolaou,
  ``A Striped Holographic Superconductor,''
  Phys.\ Rev.\ D {\bf 83}, 064009 (2011)
  [arXiv:1010.1775 [hep-th]].
\bibitem{Hutasoit:2011rd}
  J.~A.~Hutasoit, S.~Ganguli, G.~Siopsis and J.~Therrien,
  ``Strongly Coupled Striped Superconductor with Large Modulation,''
  JHEP {\bf 1202}, 086 (2012)
  [arXiv:1110.4632 [cond-mat.str-el]].
\bibitem{Ganguli:2012up}
  S.~Ganguli, J.~A.~Hutasoit and G.~Siopsis,
  ``Enhancement of Critical Temperature of a Striped Holographic Superconductor,''
  Phys.\ Rev.\ D {\bf 86}, 125005 (2012)
  [arXiv:1205.3107 [hep-th]].
\bibitem{Hutasoit:2012ib}
  J.~A.~Hutasoit, G.~Siopsis and J.~Therrien,
  ``Conductivity of Strongly Coupled Striped Superconductor,''
  arXiv:1208.2964 [hep-th].
\bibitem{Horowitz:2012ky}
  G.~T.~Horowitz, J.~E.~Santos and D.~Tong,
  ``Optical Conductivity with Holographic Lattices,''
  JHEP {\bf 1207}, 168 (2012)
  [arXiv:1204.0519 [hep-th]].
\bibitem{Horowitz:2012gs}
  G.~T.~Horowitz, J.~E.~Santos and D.~Tong,
  ``Further Evidence for Lattice-Induced Scaling,''
  JHEP {\bf 1211}, 102 (2012)
  [arXiv:1209.1098 [hep-th]].
\bibitem{Horowitz:2013jaa}
  G.~T.~Horowitz and J.~E.~Santos,
  ``General Relativity and the Cuprates,''
  arXiv:1302.6586 [hep-th].
\bibitem{Ling:2013aya}
  Y.~Ling, C.~Niu, J.~Wu, Z.~Xian and H.~Zhang,
  ``Holographic Fermionic Liquid with Lattices,''
  arXiv:1304.2128 [hep-th].
 \bibitem{aperis}
  A.~Aperis, P.~Kotetes, E.~Papantonopoulos, G.~Siopsis, P.~Skamagoulis and G.~Varelogiannis,
  ``Holographic Charge Density Waves,''
  Phys.\ Lett.\ B {\bf 702}, 181 (2011)
  [arXiv:1009.6179 [hep-th]].
\bibitem{Domokos:2007kt}
  S.~K.~Domokos and J.~A.~Harvey,
  ``Baryon number-induced Chern-Simons couplings of vector and axial-vector mesons in holographic QCD,''
  Phys.\ Rev.\ Lett.\  {\bf 99}, 141602 (2007)
  [arXiv:0704.1604 [hep-ph]].
\bibitem{ooguri} S. Nakamura, H. Ooguri, and C.-S. Park, ``Gravity Dual of Spatially Modulated
Phase," Phys. Rev. D81 (2010) 044018, arXiv:0911.0679 [hep-th].
\bibitem{ooguri1} H. Ooguri and C.-S. Park, ``Holographic End-Point of Spatially Modulated
Phase Transition," Phys. Rev. D82 (2010) 126001, arXiv:1007.3737
[hep-th].
\bibitem{Ammon:2011je}
  M.~Ammon, J.~Erdmenger, P.~Kerner and M.~Strydom,
  ``Black Hole Instability Induced by a Magnetic Field,''
  Phys.\ Lett.\ B {\bf 706}, 94 (2011)
  [arXiv:1106.4551 [hep-th]].
\bibitem{Bu:2012mq}
  Y.~-Y.~Bu, J.~Erdmenger, J.~P.~Shock and M.~Strydom,
  ``Magnetic field induced lattice ground states from holography,''
  JHEP {\bf 1303}, 165 (2013)
  [arXiv:1210.6669 [hep-th]].

\bibitem{Donos1} A. Donos and S. A. Hartnoll, ``Metal-Insulator Transition in Holography,"
arXiv:1212.2998 [hep-th].
\bibitem{h9} A. Donos, J. P. Gauntlett, J. Sonner, and B. Withers, ``Competing orders in
M-theory: superfuids, stripes and metamagnetism," JHEP 1303 (2013) 108,
arXiv:1212.0871 [hep-th].

\bibitem{h7} N. Bao, S. Harrison, S. Kachru, and S. Sachdev, ``Vortex Lattices and
Crystalline Geometries," arXiv:1303.4390 [hep-th].



\bibitem{witten} E. Witten, ``Anti-de Sitter space, thermal phase transition, and confinement in gauge theories,"
Adv. Theor. Phys. {\bf 2} (1998) 505 [arXiv:hep-th/9803131]
\bibitem{horowitz1} G. T. Horowitz and R. C. Myers,
``The AdS/CFT correspondence and a new positive energy conjecture for geneneral relativity,"
Phys. Rev. D {\bf 59} (1998) 026005 [arXiv:hep-th/9808079].

\bibitem{siopsis} G. Siopsis, J. Therrien,  ``Analytic Calculation of Properties of Holographic Superconductors,''
  JHEP {\bf 1005}, 013 (2010)
  [arXiv:1003.4275 [hep-th]].
\bibitem{cai1}R.~-G.~Cai, H.~-F.~Li and H.~-Q.~Zhang,
  ``Analytical Studies on Holographic Insulator/Superconductor Phase Transitions,''
  Phys.\ Rev.\ D {\bf 83}, 126007 (2011)
  [arXiv:1103.5568 [hep-th]].
\bibitem{pan}Q.~Pan, J.~Jing and B.~Wang,
  ``Analytical investigation of the phase transition between holographic insulator and superconductor in Gauss-Bonnet gravity,''
  JHEP {\bf 1111}, 088 (2011)
  [arXiv:1105.6153 [gr-qc]].
\bibitem{cai2}R.~-G.~Cai, L.~Li, H.~-Q.~Zhang and Y.~-L.~Zhang,
  ``Magnetic Field Effect on the Phase Transition in AdS Soliton Spacetime,''
  Phys.\ Rev.\ D {\bf 84}, 126008 (2011)
  [arXiv:1109.5885 [hep-th]].
\bibitem{pan2} Q.~Pan, J.~Jing, B.~Wang and S.~Chen,
  ``Analytical study on holographic superconductors with backreactions,''
  JHEP {\bf 1206}, 087 (2012)
  [arXiv:1205.3543 [hep-th]].
\bibitem{pan20}
X.~-H.~Ge, B.~Wang, S.~-F.~Wu and G.~-H.~Yang,
  ``Analytical study on holographic superconductors in external magnetic field,''
  JHEP {\bf 1008}, 108 (2010)
  [arXiv:1002.4901 [hep-th]].\\
X.~-H.~Ge and H.~-Q.~Leng,
  ``Analytical calculation on critical magnetic field in holographic superconductors with backreaction,''
  Prog.\ Theor.\ Phys.\  {\bf 128}, 1211 (2012)
  [arXiv:1105.4333 [hep-th]].
\bibitem{lee}
 C.~O.~Lee,
  ``The holographic superconductors in higher-dimensional AdS soliton,''
  Eur.\ Phys.\ J.\ C {\bf 72}, 2092 (2012)
  [arXiv:1202.5146 [gr-qc]].
\bibitem{zhao}Z.~Zhao, Q.~Pan and J.~Jing,
  ``Holographic insulator/superconductor phase transition with Weyl corrections,''
  Phys.\ Lett.\ B {\bf 719}, 440 (2013)
  [arXiv:1212.3062 [hep-th]].
\bibitem{gang}S. Gangopadhyay,
 ``Analytic study of properties of holographic superconductors away from the probe limit
 [arXiv:1302.1288 [hep-th] ]; \\S.~Gangopadhyay and D.~Roychowdhury,
  ``Analytic study of Gauss-Bonnet holographic superconductors in Born-Infeld electrodynamics,''
  JHEP {\bf 1205}, 156 (2012)
  [arXiv:1204.0673 [hep-th]].
\bibitem{wang} Y.~Peng, Q.~Pan and B.~Wang,
  ``Various types of phase transitions in the AdS soliton background,''
  Phys.\ Lett.\ B {\bf 699}, 383 (2011)
  [arXiv:1104.2478 [hep-th]].
\bibitem{Domokos:2013kha}
  S.~K.~Domokos, C.~Hoyos and J.~Sonnenschein,
  ``Stability conditions for spatially modulated phases,''
  arXiv:1307.3773 [hep-th].
\end{thebibliography}
\end{document}